\documentstyle[12pt]{article}
\newcommand{\bce}{\begin{center}}
\newcommand{\ece}{\end{center}}

\newcommand{\ba}{\begin{array}}
\newcommand{\ea}{\end{array}}

\def\lsim{\mathrel{\rlap{\lower4pt\hbox{\hskip1pt$\sim$}}
    \raise1pt\hbox{$<$}}}         %less than or approx. symbol
\def\gsim{\mathrel{\rlap{\lower4pt\hbox{\hskip1pt$\sim$}}
    \raise1pt\hbox{$>$}}}         %greater than or approx. symbol

\def\beq{\begin{equation}}
\def\endeq{\end{equation}}
\def\arr{\begin{eqnarray}}
\def\endarr{\end{eqnarray}}
\def\Pom{{\bf I\!P}}

  \advance\oddsidemargin  by -1in
  \advance\evensidemargin by -1in
  \advance\topmargin      by -0.10in
\makeindex

\begin{document}

\hspace*{9.65cm} {\Large \bf DFTT 71/95} \\
\hspace*{10.3cm} {\Large \bf KFA-IKP(TH)-24-95}\\
\hspace*{10.3cm} {\Large \bf hep-ph/9605231} 
\vspace*{1.50cm}

\begin{center}
{\huge \bf
Color dipole phenomenology of
diffractive electroproduction
of light vector mesons at HERA}
\vspace{0.4cm}\\
 {\large \bf
J.Nemchik$^{a,b}$, N.N.~Nikolaev$^{c,d}$, E.~Predazzi$^{a}$,
B.G.~Zakharov$^{a,d}$
\bigskip\\}
{\sl
$^{a}$Dipartimento di Fisica Teorica, Universit\` a di Torino,\\
and INFN, Sezione di Torino, I-10125, Torino, Italy
\medskip\\
$^{b}$Institute of Experimental Physics, Slovak Academy of Sciences,\\
Watsonova 47, 04353 Kosice, Slovak Republik
\medskip\\
$^{c}$IKP(Theorie), KFA J{\"u}lich, 5170 J{\"u}lich, Germany
\medskip\\
$^{d}$L. D. Landau Institute for Theoretical Physics, GSP-1,
117940, \\
ul. Kosygina 2, Moscow 117334, Russia.}
\end{center}

\vspace{0.4cm}

\begin{abstract}

We develop the color dipole phenomenology of diffractive photo-
and electroproduction $\gamma^{*}\,N\rightarrow V(V')\,N$ of
light vector mesons ($V(1S) = \phi^0, \omega^0, \rho^0$) and their radial
excitations ($V'(2S) = \phi', \omega', \rho'$). The
node of the radial wave function of the $2S$ states in conjunction
with the energy dependence of the color dipole cross section
is shown to lead to a strikingly different $Q^2$ and $\nu$ dependence of
diffractive production of the $V(1S)$ and $V'(2S)$ vector mesons.
We discuss the restoration of flavor symmetry and
universality properties
of production of different vector
mesons as a function of
 $Q^{2}+m_{V}^{2}$. The color dipole model
predictions for the $\rho^{0}$ and $\phi^{0}$ production
are in good agreement with the experimental data from the
EMC, NMC, ZEUS and H1 collaborations. We present the first
direct evaluation of the dipole cross section from these data.
\\

\end{abstract}

E-mails:\\
bgz@landau.ac.ru \\
nemchik@to.infn.it\\
kph154@aix.sp.kfa-juelich.de\\
predazzi@to.infn.it
\baselineskip0.8cm
\pagebreak

%-------------------------------------------------

%     Section 1

%------------------------

\section{Introduction}

~~~~Diffractive electroproduction of vector mesons
\beq
\gamma^{*}p\rightarrow Vp\,,~~~~~V=\rho^{0},\,\omega^{0},\,
\phi^{0},\,J/\Psi,\,\Upsilon\,
\label{eq:1.1}
\endeq
at high energy $\nu$
offers a unique possibility of studying the pomeron exchange
at high energies
\cite{DL,KZ91,Ryskin,KNNZ93,KNNZ94,NNZscan,Brodsky,
Forshaw}. Particularly important is the observation that
the transverse size of the photon shrinks with the increase
of its virtuality $Q^{2}$. This property can
conveniently be quantified in the mixed $({\bf{r}},z)$
lightcone technique \cite{NZ91,NZ94}, in which the high energy
hadrons and photons are described as systems of color dipoles
with the transverse size ${\bf{r}}$ frozen during
the interaction process.
Interaction of color dipoles with the target nucleon is
quantified by the color dipole cross section
$\sigma(\nu,r)$ whose evolution
with the energy $\nu$ is described by the generalized BFKL
equation \cite{NZ94,NZZ94} (for a related approach
see also \cite{Mueller}). The shrinkage of the photon with
$Q^{2}$ together with the small-size
behavior of the dipole cross section ($\sim r^{2}$)
leads to what has come to be known as a
{\it scanning phenomenon} \cite{NNN92,KNNZ93,KNNZ94,NNZscan}: the
$V(1S)$ vector meson production amplitude is dominated by
the contribution from the dipole cross
section at the dipole size $r\sim r_{S}$, where
$r_{S}$ is the scanning radius
\beq
r_{S} \approx {A \over \sqrt{m_{V}^{2}+Q^{2}}}\, .
\label{eq:1.2}
\endeq
This scanning property makes the vector meson production an
ideal laboratory for testing the generalized BFKL dynamics [gBFKL hereafter].
At large $Q^{2}$ and/or for heavy vector mesons,
the amplitude of reaction (\ref{eq:1.1}) becomes short-distance
dominated and is perturbatively calculable in terms of the
short-distance behavior of the vector mesons wave function. However,
the asymptotic short distance formulas \cite{Ryskin,Brodsky}
are not yet applicable at the moderate $Q^{2} \lsim 20$\,GeV$^{2}$
of interest in the present fixed target and HERA experiments
where the scanning radius
$r_{S}$ is still large due to a large scale parameter
$A\approx 6$ in (\ref{eq:1.2}) as derived in \cite{NNZscan}. For this
reason, the onset of the short-distance dominance is very slow
and there emerges a unique possibility of studying the transition
between the soft and hard interaction regimes in a well controlled
manner. Furthermore, the scanning phenomenon allows to directly
test the steeper
subasymptotic energy dependence of the dipole cross section at
smaller dipole size $r$, which is one of
interesting consequences of the color dipole gBFKL dynamics
\cite{NZZ94,NZBFKL}.

The scanning phenomenon has particularly interesting
implications for the
 diffractive production of the $2S$ radially excited vector
mesons
\beq
\gamma^{*}p\rightarrow V'p\,,~~~~~V'(2S)=\rho',\,\omega',\,
\phi',\,\Psi',\,\Upsilon'\,..
\label{eq:1.3}
\endeq
Here one encounters the node effect: a tricky and strong cancellation
between the large and the small size contributions to the production
amplitude i.e., those above and below the node position
$r_{n}$ in the $2S$ radial wave function \cite{KZ91,NNN92,NNZanom}
respectively. The node effect
is the only dynamical mechanism that gives a strong natural
suppression of the photoproduction of excited vector mesons $V'(2S)$
vs. $V(1S)$ mesons.
For instance, it correctly
predicted \cite{KZ91,NNN92} the
strong suppression of real photoproduction of the $\Psi'$ compared
to the $J/\Psi$ observed in the NMC experiment
\cite{NMCPsi'} and confirmed recently in the
high statistics E687 experiment \cite{E687Psi'}. In anticipation
of the new experimental data on real and virtual $V'$
photoproduction from HERA, it is important to further
explore the salient features of the node effect in the
framework of the color dipole gBFKL dynamics. At
moderate $Q^{2}$, the scanning radius
$r_{S}$ is comparable to $r_{n}$. First, for this reason
even a slight variation of
$r_{S}$ with $Q^{2}$ leads to a strong change of the cancellation
pattern in the $V'(2S)$ production amplitude and to an anomalous
$Q^{2}$ dependence for the electroproduction of the radially excited
vector mesons \cite{KZ91,NNN92,NNZanom}. Second,
the cancellation pattern is sensitive also to the dipole-size
dependence
of the color dipole
cross section
$\sigma(\nu,r)$ which in the gBFKL
dynamics changes which energy $\nu$ leading to an
anomalous energy dependence for producing the $V'(2S)$
vector mesons as compared to a smooth energy dependence for
the $V(1S)$ ground state vector mesons. This anomalous
$Q^{2}$ and energy dependence
of the $V'(2S)$ production offers a unique signature of
the $2S$ radial excitation vs. the D-wave state.
Third, at very small $Q^{2}$, the $V'(2S)$ production amplitude can be of
opposite sign with respect to that of the $V(1S)$ production amplitude
(the overcompensation scenario of ref. \cite{NNZanom})
to then conform to the same
sign at larger $Q^{2}$ (the undercompensation scenario
of ref. \cite{NNZanom}). Here we wish to emphasize that
the relative sign of the $V'$ and $V$ production amplitudes
is experimentally measurable using the so-called S\"oding-Pumplin
effect (\cite{Soding,Pumplin}, see also \cite{Abe}).

In this paper we develop the color dipole phenomenology of
diffractive photo- and electroproduction of the $1S$ ground
state and of the $2S$ radially excited vector mesons. As stated
above, for a large scanning radius, the large
distance contribution to the production amplitude is not yet
negligibly small in the so far experimentally studied region
of $Q^{2}$, in particular in the $2S$ meson production.
In this paper we show that
the $Q^{2}$ and energy dependence
of the diffractive production of vector mesons offers a unique
possibility of studying how the color dipole cross section
changes from the large nonperturbative to the small perturbative
dipole size. The problem can be attacked both ways.
First, we present detailed predictions using the
color dipole cross section \cite{NZHera,NNZscan}, which gives
a very good quantitative description of the proton structure function
from very small to large $Q^{2}$.
Second,
we can invert the problem and evaluate the color dipole cross
section from the corresponding experimental data. Such an
evaluation of the dipole cross section is presented here for
the first time.

The paper is organized as follows. In section 2 we formulate
the color dipole factorization for vector meson production
amplitudes. In section 3 we present our numerical results.
We find good agreement with the experimental data from the
fixed target and HERA collider experiments.
The subject of section 4 is the anomalous $Q^{2}$ and
energy dependence of electroproduction of $2S$ radially
excited vector mesons. In section 5 we discuss the scaling
relations between production cross sections for different
vector mesons and the restoration of flavor symmetry in
the variable $Q^{2}+m_{V}^{2}$.
We comment on how the scanning phenomenon
enables a direct comparison of the spatial wave functions of the
$\rho^{0}$ and $\omega^{0}$ mesons. The first evaluation
of the dipole cross
section from the experimental data is presented in section 6.
In section 7 we summarize
our main results and conclusions. In the Appendix we describe
the lightcone parameterization of wave functions of $V(1S)$ and
$V(2S)$ vector mesons used in our analysis.

%----------------------------------------------------

%            Section 2

%----------------

\section{Color dipole factorization for vector meson production}

The Fock state expansion for the relativistic meson starts
with the quark-antiquark state which can be considered as
a color dipole. The relevant variables
are the dipole moment
$\bf{r}$ which is the transverse separation
(with respect to the collision axis) of the quark and
antiquark and $z$ - the fraction of the lightcone momentum of the
meson
carried by a quark. The interaction of the relativistic
color dipole with the
target nucleon is described by the energy dependent color
dipole cross section $\sigma(\nu,r)$.
The many gluon contributions of
higher Fock states $q\bar{q}g...$ become very important at high
energy $\nu$. The crucial point is that in the leading log${1\over x}$
the
effect of higher Fock states can be reabsorbed into the energy
dependence of $\sigma(\nu,r)$, which satisfies the generalized
BFKL equation (\cite{NZ94,NZZ94}). The flavor blind
(one should really
say {\it flavor tasteless}) dipole cross
section unifies the description of various diffractive processes.
To apply the color dipole
formalism to deep inelastic and quarkonium scattering and
diffractive production of vector mesons one needs the
probability amplitudes $\Psi_{\gamma^{*}}(\vec{r},z)$ and
$\Psi_{V}(\vec{r},z)$
to find the color dipole of size $r$
in the photon and quarkonium (vector meson), respectively.
The color dipole distribution in (virtual) photons was derived in
\cite{NZ91,NZ94}.
In terms of these probability amplitudes, the imaginary part
of the
virtual photoproduction of vector mesons in the forward direction
($t=0$) reads
\beq
{\rm Im}{\cal M}=\langle V|\sigma(\nu,r)|\gamma^{*}\rangle
=\int_{0}^{1} dz \int d^{2}{\bf{r}}\,\sigma(\nu,r)
\Psi_V^{*}({\bf{r}},z)
\Psi_{\gamma^{*}}({\bf{r}},z) \,
\label{eq:2.2}
\endeq
whose normalization is
$
\left.{d\sigma/ dt}\right|_{t=0}={|{\cal M}|^{2}/ 16\pi}.
$
For small size heavy quarkonium the probability amplitude
$\Psi_{V}({\bf{r}},z)$ can
safely be identified with the constituent quark-antiquark
quarkonium wave function.
The color dipole factorization (\ref{eq:2.2}) takes
advantage of
the diagonalization of the scattering
matrix in the $({\bf{r}},z)$ representation, which
clearly holds even when the dipole
size ${\bf{r}}$ is large,
i.e. beyond the perturbative region of short
distances. Due to this property and to the fact
that in leading log${1\over x}$  the effect of higher Fock states
is reabsorbed in the energy dependence of the dipole cross
section $\sigma(\nu,r)$, as a starting approximation we can
identify the probability amplitude $\Psi_{V}(\vec{r},z)$
for large size dipoles in light vector mesons
with  the constituent quark wave function
of the meson.
This provides a viable
phenomenology of diffractive scattering which is purely
perturbative for small size mesons and/or
large $Q^{2}$ and small scanning radius
$r_{S}$ and allows a sensible
interpolation between soft interactions for large dipoles and
hard perturbative interactions of small dipoles.
For light quarkonia and small $Q^{2}$, this
implies the assumption that small-size constituent quarks are
the relevant degrees of freedom and the spatial separation
of constituent quarks is a major dynamical variable in the
scattering process.
%kolya begin
{%\bf
\footnote{
%\bf
See also earlier works on the color dipole analysis of
hadronic diffractive interactions which
used constituent quark wave functions for
the color dipole distribution amplitudes \cite{Dipole}.}
The large-$r$ contribution to the production amplitude (\ref{eq:3.1})
depends on the both dipole cross section for large-size dipoles
and the amplitudes of distribution of large-size color dipoles
and/or the nonperturbative wave functions of light vector mesons
at large $r$, both of which are poorly known at the moment. Still,
testing the predictions from such a minimal
model is interesting for its own sake and can shed a light
on the transition between the soft and
hard scattering regimes which is still far from understood. An
analysis of sensitivity to models of the nonperturbative wave
functions of vector mesons and of how one can disentangle the
effects of large $r$ behavior of the wave function and of the
dipole cross section, goes beyond the scope of the present
exploratory study.
}
%kolya end

The energy dependence of the dipole cross section is quantified
in terms of the dimensionless
rapidity $\xi$, which in deep inelastic scattering
equals $\xi = \log{1\over x}$.
%kolya  begin
{%\bf
Considerations of intermediate
masses in diagrams for exclusive production of vector
mesons show that to the considered leading log${1\over x}$
approximation one must take $\xi=\log{1\over x_{eff}}$, where
%kolya end
}
\beq
x_{eff}= {Q^{2}+m_{V}^{2}\over 2m_{p}\nu}
\, ,
\label{eq:2.3}
\endeq
and  $m_{V}$ is a mass of the vector meson. The pomeron exchange
dominance holds when the Regge parameter is large,
% ----------------------------------------------------------
\beq
\omega={1\over x_{eff}}={2m_{p}\nu \over (Q^{2}+m_{V}^{2})}\gg 1\,.
\label{eq:3.1}
\endeq
% -----------------------------------------------------------
%
 Hereafter we
write the amplitudes in terms of $\sigma(x_{eff},r)$.
The spin independence
of the dipole cross section $\sigma(x_{eff},r)$ in (\ref{eq:2.2})
leads to the $s$-channel helicity conservation: the transversely
polarized photons produce transversely polarized vector mesons and
the longitudinally  polarized vector mesons are produced by
longitudinal (to be more precise, the scalar one) photons. More explicitly,
the form of the forward production amplitudes for the
transversely (T) and the longitudinally (L) polarized vector mesons
in terms of the lightcone radial wave function $\phi(r,z)$
of the $q\bar{q}$ Fock state of the vector meson reads \cite{NNZscan}
\arr
{\rm Im}{\cal M}_{T}(x_{eff},Q^{2})=
{N_{c}C_{V}\sqrt{4\pi\alpha_{em}} \over (2\pi)^{2}}
\cdot~~~~~~~~~~~~~~~~~~~~~~~~~~~~~~~~~
\nonumber \\
\cdot \int d^{2}{\bf{r}} \sigma(x_{eff},r)
\int_{0}^{1}{dz \over z(1-z)}\left\{
m_{q}^{2}
K_{0}(\varepsilon r)
\phi(r,z)-
[z^{2}+(1-z)^{2}]\varepsilon K_{1}(\varepsilon r)\partial_{r}
\phi(r,z)\right\}\nonumber \\
 =
{1 \over (m_{V}^{2}+Q^{2})^{2}}
\int {dr^{2} \over r^{2}} {\sigma(x_{eff},r) \over r^{2}}
W_{T}(Q^{2},r^{2})
\label{eq:2.4}
\endarr
\arr
{\rm Im}{\cal M}_{L}(x_{eff},Q^{2})=
{N_{c}C_{V}\sqrt{4\pi\alpha_{em}} \over (2\pi)^{2}}
{2\sqrt{Q^{2}} \over m_{V}}
\cdot~~~~~~~~~~~~~~~~~~~~~~~~~~~~~~~~~
 \nonumber \\
\cdot \int d^{2}{\bf{r}} \sigma(x_{eff},r)
\int_{0}^{1}dz \left\{
[m_{q}^{2}+z(1-z)m_{V}^{2}]
K_{0}(\varepsilon r)
\phi(r,z)-
\varepsilon K_{1}(\varepsilon r)\partial_{r}
\phi(r,z)\right\}\nonumber \\
 =
{1 \over (m_{V}^{2}+Q^{2})^{2}}
{2\sqrt{Q^{2}} \over m_{V}}
\int {dr^{2} \over r^{2}} {\sigma(x_{eff},r) \over r^{2}}
W_{L}(Q^{2},r^{2})
\label{eq:2.5}
\endarr
where
\beq
\varepsilon^{2} = m_{q}^{2}+z(1-z)Q^{2}\,,
\label{eq:2.6}
\endeq
$\alpha_{em}$ is the fine structure
constant, $N_{c}=3$ is the number of colors,
$C_{V}={1\over \sqrt{2}},\,{1\over 3\sqrt{2}},\,{1\over 3},\,
{2\over 3}~~$ for
$\rho^{0},\,\omega^{0},\,\phi^{0},\, J/\Psi$ production,
respectively and
$K_{0,1}(x)$ are the modified Bessel functions.
The detailed discussion and parameterization of $\phi(r,z)$
is given in the Appendix, here we only mention that
the form of $\phi(r,z)$ we use has the hard-QCD driven
short distance behavior and
gives the electromagnetic form
factor of mesons which has the correct QCD asymptotic behavior.
At large $r$ we follow the conventional spectroscopic models
\cite{Potential}
and constrain the parameters of the wave functions by the
widths of the leptonic decays $V,V'\rightarrow e^{+}e^{-}$, the
radii of the vector mesons and the 2S-1S mass splitting.
The terms $\propto \phi(r,z)K_{0}(\varepsilon r)$ and
$\propto \partial_{r}\phi(r,z) \varepsilon
K_{1}(\varepsilon r)$, i.e., $ \partial_{r}\phi(r,z)
\partial_{r}K_{0}(\varepsilon r)$,
in the integrands
of (\ref{eq:2.4}) and (\ref{eq:2.5})
%kolya begin
{%\bf
derive from the helicity conserving and helicity
nonconserving transitions
$\gamma^{*} \rightarrow q\bar{q}$ and $V\rightarrow q\bar{q}$
in the $A_{\mu}\Psi \gamma_{\mu} \Psi$ and
$V_{\mu}\Psi \gamma_{\mu} \Psi$ vertices
(see Bjorken et al. \cite{Bjorken} and \cite{NZ91}, the technique of
calculation of traces in the spinorial representation of the relevant
Feynman amplitudes is given in \cite{NZ91} and need not be repeated
here;
for the related Melosh
transformation analysis see the recent Ref. \cite{Gevorkyan}).
}
%kolya end
The latter are the relativistic corrections,
for the heavy quarkonium
the nonrelativistic
approximation \cite{KZ91} has a rather high accuracy,
the relativistic corrections become important only
at large $Q^{2}$ and for the production of
light vector mesons.
%kolya begin
{%\bf
%(see also the comparison with
%the approach of Ref.~\cite{Brodsky} in Section 6).
}
%kolya end
Eqs. (\ref{eq:2.4}),(\ref{eq:2.5}) give the imaginary part of the production
amplitudes; one can easily include small corrections for the real
part by the substitution \cite{GribMig},
\beq
\sigma(x_{eff},r) \Longrightarrow \left(1-i
\cdot\frac{\pi}{2}\cdot\frac{\partial}
{\partial\,\log\,x_{eff}} \right)\sigma(x_{eff},r) =
\Biggl[1-i\cdot\alpha_{V}(x_{eff},r)\Biggr]\sigma(x_{eff},r)
\label{eq:2.7}
\endeq
For brevity, in the subsequent discussion
we suppress the real part of the production
amplitude; it is consistently included in all the numerical
calculations.

The color dipole cross section is flavor blind. The only
kinematical sensitivity to the vector meson produced comes via the
rapidity variable, see Eq.~(\ref{eq:2.3}).
For small $r$,
in the leading log${1\over x}$
and leading log${1\over r^{2}}$,
i.e., leading log$Q^{2}$,
  the dipole cross section can be related \cite{Barone} to the
gluon structure function $G(x,\bar{Q}^2)$ of the target nucleon
through
\beq
\sigma(x,r) =
\frac{\pi^2}{3}r^2\alpha_s(r)G(x,\bar{Q}^2) \, ,
\label{eq:2.8}
\endeq
%Enrico would it not be better to define $q^2$?
where the gluon structure function enters at
the factorization scale
$\bar{Q}^2 \sim {B\over r^2}$ (for the origin
of the large scale factor $B\sim 10$, see
\cite{NZglue}).
The integrands of (\ref{eq:2.4}),(\ref{eq:2.5}) are smooth at
small $r$ and decrease exponentially at $r > 1/\epsilon$ due to
the exponential decrease of the modified Bessel functions.
Together with the $\propto r^{2}$ behavior of the color
dipole cross section (\ref{eq:2.8}), this implies that the
amplitudes (\ref{eq:2.4}),(\ref{eq:2.5}) receive their dominant
contribution from $r \approx r_{S}$. (Eq.~(\ref{eq:1.2})
assumes that the scanning radius $r_{S}$ is substantially
smaller than the radius $R_{V}$ of the vector meson.)
Then, a
simple evaluation gives \cite{KNNZ94}
\beq
{\rm Im}
{\cal M}_{T} \propto r_{S}^{2}\sigma(x_{eff},r_{S}) \propto
{1 \over Q^2+m_{V}^{2}}\sigma(x_{eff},r_{S}) \propto
 {1\over (Q^{2}+m_{V}^{2})^{2} }
\label{eq:2.9}
\endeq
and
\beq
{\rm Im}
{\cal M}_{L} \approx {\sqrt{Q^{2}}\over m_{V}}{\cal M}_{T}
 \propto
{\sqrt{Q^{2}}\over m_{V}}
 r_{S}^{2}\sigma(x_{eff},r_{S})
 \propto
{\sqrt{Q^{2}}\over m_{V}}
 {1\over (Q^{2}+m_{V}^{2})^{2}}
\label{eq:2.10}
\endeq
respectively.
\footnote{
Unless otherwise specified, for each flavor, $m_{V}$ will
always be the mass of the ground state 1S vector meson.
}
%kolya begin
{%\bf
The prediction of the dominance of the longitudinal
cross section at large $Q^{2}$ is shared by all the models of
diffractive leptoproduction, starting with the vector dominance
model (\cite{Ryskin,KNNZ94,Brodsky}, for the excellent review
of early works on photo- and electroproduction of vector
mesons and on vector dominance model
see Bauer et al. \cite{Bauer}) and is confirmed by all
the experiments on leptoproduction
of the $\rho^{0}$ at large $Q^{2}$ \cite{E665rho,NMCfirho,ZEUSrhoQ2}.
}
%kolya end
The first factor $\propto r_{S}^{2} \propto 1/(Q^{2}+m_{V}^{2})$
in (\ref{eq:2.9}) comes from
the overlap of wave function of the shrinking photon and that
of the vector meson.
The familiar vector dominance model (VDM)
prediction is  $M_T \propto {1\over (m_V^2+Q^2)}\sigma_{tot}(\rho N)$,
whereas in our QCD approach a small
$\sigma(x_{eff},r_{S})\propto r_{S}^{2} \propto 1/(Q^{2}+m_{V}^{2})$
enters instead of $\sigma_{tot}(VN)$.
In (\ref{eq:2.9}),(\ref{eq:2.10}) we show only the leading
$Q^{2}$ dependence, suppressing the phenomenologically
important departure form the law $\sigma(x,r) \propto r^2$,
whose large
$Q^{2}$ dependence  can be related to
scaling violations in the gluon density
(see (\ref{eq:2.8}) and the discussion below).
%kolya begin
{%\bf
We recall that
the shrinkage of the virtual photons and/or the
decrease of the scanning radius $r_{S}$ with $Q^{2}$ is the
origin of color transparency effects in diffractive
leptoproduction of vector mesons off nuclei
\cite{KZ91,KNNZ93,KNNZ94,BrodskyMueller}. The important
confirmation of the quantitative
predictions \cite{KNNZ93,KNNZ94}
of color transparency effects based on the same technique
as used here came from the E665 experiment \cite{E665}.

More accurate analysis  of the scanning phenomenon can be performed
in terms of the weight functions $W_{T,L}(Q^{2},r^{2})$ which are
sharply peaked at $r\approx A_{T,L}/\sqrt{Q^{2}+m_{V}^{2}}$,
in the relevant variable $\log r$ the width of the peak in
$W_{L}(Q^{2},r^{2})$ at half maximum equals $\Delta \log r
\approx 1.2$ for the $J/\Psi$ production and $\Delta \log r
\approx 1.3$ for the $\rho^{0}$ production and varies little
with $Q^{2}$  \cite{NNZscan}.
}
%kolya end
The values of the scale parameter $A_{T,L}$ turn out to be
close to $A \sim 6$, which follows from $r_{S}=3/\varepsilon$ with
the nonrelativistic choice $z=0.5$; in general $A_{T,L} \geq 6$
and increases slowly with $Q^2$.
This $Q^{2}$ dependence of $A_{T,L}$ comes from
the large-size asymmetric $q\bar{q}$ configurations
when, for instance, the antiquark and the quark
in the photon and in the vector mesons carry a very large
and a very small fraction of the meson momentum
respectively (or the other way around).
A comparison
of the integrands in eqs. (\ref{eq:2.4}) and (\ref{eq:2.5}) shows that
the latter contains an extra factor $z(1-z)$
which makes considerably smaller the
contribution from asymmetric configurations to the longitudinal
meson production. For completeness,
we quote the results of \cite{NNZscan}:
$A_{T,L}(J/\Psi;Q^{2}=0)\approx 6,~
A_{T,L}(J/\Psi;Q^2 = 100\,{\rm GeV}^2) \approx 7,~
A_{L}(\rho^0;Q^{2}=0) \approx 6.5,~
A_{L}(\rho^0;Q^2 = 100\,{\rm GeV}^2) \approx 10,~
A_{T}(\rho^0;Q^{2}=0) \approx 7,~
A_{T}(\rho^0,Q^2 = 100\,{\rm GeV}^2)\approx 12.$

An alternative formulation of the slow onset of the purely
perturbative regime can be seen as follows:
at very large $Q^{2}$  when the scanning radius is very small,
the dipole cross section $\sigma(x_{eff},r)$ and the
vector meson production amplitudes are proportional to
the gluon density $G(x_{eff},\bar{Q}^{2})$
at the factorization scale
$\bar{Q}^{2}=\tau (Q^{2}+m_{V}^{2})$
%kolya begin
{%\bf
(see also Refs. \cite{Ryskin,Brodsky} which use a different
technique of the momentum-space wave functions, related to
the color dipole factorization by the Fourier-Bessel transform;
the detailed comparison with the work of Brodsky
et al. \cite{Brodsky} will be presented below in Section 6).
}
%kolya end
The large values
of $A_{T,L}$ previously quoted, reflect into
very small values of $\tau$ \cite{NNZscan}:
in the interesting region of $Q^{2}\gsim 10\,GeV^{2}$
one finds
$\tau_{T,L}(J/\Psi) \approx 0.2$, $\tau_{L}(\rho^{0})\approx 0.15$ and
$\tau_{T}(\rho^{0}) \approx $0.07-0.1, which is different
and substantially smaller than the values
 $\tau=0.25$ suggested in \cite{Ryskin} and $\tau = 1$ suggested
in \cite{Brodsky}. Very large $Q^{2}$ values are needed
for reaching the perturbatively large $\bar{Q}^{2}$ and for
the applicability of
the pQCD relationship (\ref{eq:2.8}).

Consequently, for the domain presently under
experimental study, $Q^{2}+m_{V}^{2} \lsim
$10-20\,GeV$^{2}$, the production amplitudes
receive substantial
 contribution from semiperturbative and nonperturbative $r$.
In \cite{NZHera,NNZscan} this contribution was modeled by
the energy independent soft cross section $\sigma^{(npt)}(r)$.
The particular form of this cross section successfully
predicted \cite{NZHera} the proton structure function at very
small $Q^{2}$ recently measured by the E665 collaboration
\cite{E665lowQ2}
and also gave a good description of real
photoabsorption \cite{NNZscan}.
As an example, in Fig.~1
we present an evaluation of the vector meson-nucleon
total cross section
%-----------------------------------------------
\arr
\sigma_{tot}(VN)
 = {N_{c} \over 2\pi}\int_{0}^{1}{ dz \over z^{2}(1-z)^{2} }
\int d^{2}{\bf{r}}
~\left\{m_{q}^{2} \phi(r,z)^{2}+[z^{2} +(1-z)^{2}]
[\partial_{r}\phi(r,z)]^2 \right\}  \sigma(x_{eff},r) \, .
\label{eq:2.11}
\endarr
%------------------------------------------------
The total cross section $\sigma_{tot}(\rho^{0}N)$ so found, is close to
$\sigma_{tot}(\pi N)$, and the rise of
$\sigma_{tot}(VN)$
with the c.m.s energy $W$ is consistent with the observed trend of the
hadronic total cross sections \cite{DLRegge}.
In the color dipole picture the smaller
values of $\sigma_{tot}(\phi N)$ and $\sigma_{tot}(\phi' N)$
derive from  the smaller radius of the $s\bar{s}$ quarkonium.
In the simple model
\cite{NZHera,NNZscan} the rise of $\sigma_{tot}(VN)$ is entirely
due to the gBFKL rise of the perturbative component
$\sigma^{(pt)}(x_{eff},r)$ of the dipole cross section. The rate of
rise is small for two reasons:
i) at moderate energy, $\sigma^{(pt)}(x_{eff},r)$
at large $r$ is much smaller than the
soft cross section $\sigma^{(npt)}(r)$,
ii) at large $r$ the subasymptotic effective intercept of the gBFKL
pomeron is small \cite{NZZ94,NZBFKL}. The detailed description
of the dipole cross section used in the present analysis is
given in \cite{NZHera,NNZscan} and will not be repeated here.
It is partly shown below in Fig.~16. The reason why
we focus here on this particular model
is that its success in phenomenological applications makes it
a realistic tool for the interpolation between
soft and hard
scattering regions.
Once the vector mesons wave functions are
fixed from their spectroscopic and decay properties, all the
results for diffractive real and virtual photoproduction of
vector mesons to be reported here do not contain any adjustable
parameters.

%------------------------------------------

%         Section 3

%-----------------

\section{Diffractive
$\rho^{0}$ and $\phi^{0}$
production: predictions and comparison with experiment}

The most interesting prediction from the color dipole dynamics is
a rapid decrease of production amplitudes
(\ref{eq:2.9}),(\ref{eq:2.10}) at large $Q^{2}$.
The broadest region of $Q^{2}$ was covered in the recent NMC
experiment \cite{NMCfirho} where special care was taken
to minimize the inelastic production background which plagued
earlier data on $\rho^{0}$ and $J/\Psi$ production.
In Fig.~2 we compare our predictions for $\rho^{0}$ and
$\phi^{0}$ production with the NMC data and the data from the
HERA experiments \cite{ZEUSrhoQ2,H1rho}.
Shown here is the observed polarization-unseparated
cross section $\sigma(\gamma^{*}\rightarrow V)=
\sigma_{T}(\gamma^{*}\rightarrow V)+
\epsilon \sigma_{L}(\gamma^{*}\rightarrow V)$ for the
value of the longitudinal polarization
$\epsilon$ of the virtual photon taken from the corresponding
experimental paper (typically, $\epsilon \sim 1$).
The quantity which is best predicted theoretically is
$d\sigma(\gamma^{*}\rightarrow V)/dt|_{t=0}$; in
our evaluations of the total
production cross section $\sigma(\gamma^{*}\rightarrow V)=
B(\gamma^{*}\rightarrow V)
d\sigma(\gamma^{*}\rightarrow V)/dt|_{t=0}$
we use the diffraction
slope $B(\gamma^{*}\rightarrow V)$
given in the corresponding experimental paper.

Eqs. (\ref{eq:2.4}),(\ref{eq:2.5}) describe the pure
pomeron exchange contribution to the production amplitude.
While at HERA energies secondary Reggeon exchanges
can be neglected since the Regge parameter $\omega$ is a very large,
at the lower energy of the NMC experiment,
$\langle \nu \rangle=$(90-140)\,GeV,
the Regge parameter $\omega$
is small and non-vacuum Reggeon exchange cannot be neglected.
The fit to $\sigma_{tot}(\gamma p)$ can, for instance,
be cast in the form
%
% ------------------------------------------------------------
\beq
\sigma_{tot}(\gamma p) = \sigma_{\Pom}(\gamma p)\cdot
\left(1 + \frac{A}{\omega^{\Delta}}\right)
\label{eq:3.2}
\endeq
% ------------------------------------------------------------
%
where the term $A/\omega^{\Delta}$ in the factor
$f = 1 + A/\omega^{\Delta}$ represents the non-vacuum
Reggeon exchange contribution.
The Donnachie-Landshoff fit gives
$A = 2.332$ and $\Delta = 0.533$ \cite{DLRegge}.
We do not know how large this non-vacuum contribution
to $\rho^{0}$ production is at large $Q^{2}$;
for a crude estimation we assume the Reggeon/pomeron
ratio to scale with $\omega$,
which is not inconsistent with the known decomposition
of the proton structure function into the valence
(non-vacuum Reggeon) and sea (pomeron) contributions.
Then, for the NMC kinematics we find $f = 1.25$
at $\omega \simeq 70, Q^{2} = 3$\,GeV$^{2}$ and
$f = 1.8$
at $\omega \simeq 9, Q^{2} = 20$\,GeV$^{2}$.
This departure of $f$ from unity provides
a conservative scale
for the theoretical uncertainties at moderate values of $\omega$.
Anyway, the $Q^{2}$ dependence of the Reggeon correction  factor
$f$  is weak
compared with the very rapid variations of ${\cal M}_{T}$ and
${\cal M}_{L}$ with $Q^{2}$.
The correction for the secondary exchanges,
$\sigma(\gamma^{*} \rightarrow \rho^{0}) =
f^{2} \sigma_{\Pom}(\gamma^{*} \rightarrow \rho^{0})$,
brings the theory to a better agreement with the NMC data.
The dipole cross section of \cite{NZHera,NNZscan} correctly
describes the variation of the $\rho^{0}$ production cross
section by 3 orders in magnitude from $Q^{2}=0$ to
$Q^{2}=16.5$\,GeV$^{2}$.
For $\phi^{0}$ production, $f\equiv 1$ due to the Zweig rule and
the pure pomeron contribution correctly reproduces
the magnitude of $\sigma(\gamma^{*}\rightarrow \phi^{0})$
and its variation by nearly three
orders in the magnitude from $Q^{2}=0$
to $Q^{2}=11.3$\,GeV$^{2}$.

The specific prediction from the gBFKL dynamics
is a steeper subasymptotic growth with energy
of the dipole cross
section $\sigma(\nu,r)$ at smaller dipole size $r$,
which by virtue of the scanning phenomenon
translates into a steeper rise of
$\sigma(\gamma^{*}\rightarrow V)$ at higher $Q^{2}$
and/or for heavy quarkonia.
This consequence of the color dipole dynamics was first
explored in \cite{NNZscan}; the
$\rho^{0}$ wave function parameters used in \cite{NNZscan}
are slightly different from those used here but
the difference in $\sigma(\gamma^{*}\rightarrow \rho^{0})$
is marginal.
The agreement of our high-energy results with the HERA data is good
for both $Q^{2}=0$ (Fig.~3) and large $Q^{2}$
(Fig.~2) and confirms the growth
of the dipole cross section with energy expected from the gBFKL
dynamics.

The above
high-$Q^{2}$ data are dominated by the longitudinal
cross section; real photoproduction ($Q^{2}=0$) measures
the purely transverse cross section. In Fig.~3 we present our results
with and without secondary Reggeon corrections
($d\sigma_{\Pom}(\gamma\rightarrow \rho^{0})/dt|_{t=0}$ and
$d\sigma(\gamma\rightarrow \rho^{0})/dt|_{t=0}=
f^{2}d\sigma_{\Pom}(\gamma\rightarrow \rho^{0})/dt|_{t=0}$
respectively) as a
function of energy.
The Reggeon correction factor $f^{2}$
brings the theory to a better agreement with the low energy
$\rho^{0}$ production
data \cite{Rholownu}.
Real photoproduction of $\rho^{0}$ is dominated by
the soft contribution, the growth of the
production
cross section is driven by the rising gBFKL component
of the dipole cross section.
Our predictions for high energy
agree well with the recent
ZEUS data \cite{ZEUSrho94,ZEUSrho95}.
The
$\phi^{0}$ production is pomeron dominated which implies
$f\equiv 1$. We find good agreement with the fixed target
\cite{Philownu}
and ZEUS \cite{ZEUSphi}
data on real photoproduction of the $\phi^{0}$,
although the error bars are large (Fig.~4).
Because in $\phi^{0}$ photoproduction the relevant
dipole sizes are smaller than in the $\rho^{0}$ case,
(see the radii of
$\rho^{0}$ and $\phi^{0}$ in Table 1), we predict a steep energy
dependence of the $\phi^{0}$ production forward cross section:
$d\sigma(\gamma \rightarrow \phi^{0})/dt|_{t=0}$
is predicted to grow by a
factor $\approx 2.5$
from $3.75\,\mu b/GeV^2$ at $\nu = 175\,GeV$, i.e.,
$W=18\,GeV$, up to $\sim 8.84\,\mu b/GeV^2$
at $W = 170\,GeV$ at HERA. At $W=70$\,GeV we have
$\sigma(\gamma\rightarrow \phi^{0})=0.87 \mu b$
which agrees with the first ZEUS measurement
$\sigma(\gamma\rightarrow\phi^{0})=0.95 \pm 0.33$\,$\mu b$
\cite{ZEUSphi}.
More detailed predictions for the energy and $Q^{2}$ dependence
of $d\sigma(\gamma^{*}\rightarrow V)/dt|_{t=0}$
are presented in Fig.~5 and clearly show a
steeper rise with energy at larger $Q^{2}$ (see also \cite{NNZscan}).

In Fig.~6 we show our  predictions for
%
% --------------------------------------------------------------
\beq
R_{LT}={m_{V}^{2} \over Q^{2}}{d\sigma_{L}(\gamma^{*}\rightarrow V)
\over d\sigma_{T}(\gamma^{*}\rightarrow V)}\,.
\label{eq:3.3}
\endeq
% -------------------------------------------------------------
%
The steady decrease of $R_{LT}$ with $Q^{2}$
which implies a diminution of the dominance of the longitudinal cross
section is a very specific prediction of the color dipole
approach. It follows from
a larger contribution from large size dipoles to the
production amplitude for the
transversely polarized vector mesons and larger value of
the average scanning radius, i.e.,
$A_{T} \gsim A_{L}$ \cite{NNZscan}.
This prediction can be checked
with the higher precision data from HERA;
the available experimental data
\cite{E665rho,NMCfirho,ZEUSrhoQ2}
agree with $R_{LT} < 1$ but
have still large error bars.

The $Q^{2}$ dependence of the observed polarization-unseparated
cross section depends on
the longitudinal
polarization $\epsilon$
of the virtual photon.
To a
crude approximation the color dipole dynamics predicts
%
% ----------------------------------------------------------
\beq
\sigma(\gamma^{*}\rightarrow V)=
\sigma_{T}(\gamma^{*}\rightarrow V)+
\epsilon \sigma_{L}(\gamma^{*}\rightarrow V) \propto
{1 \over (Q^{2}+m_{V}^{2})^{4}}
\left(1+\epsilon R_{LT} {Q^{2}\over m_{V}^{2}}\right)
\label{eq:3.4}
\endeq
% ---------------------------------------------------------
%
If one approximates (\ref{eq:3.4}) by the
$(Q^{2}+m_{V}^{2})^{-n}$ behavior, one finds
$n \sim 3$ vs. $n\sim 1$ in the naive VDM.
In (\ref{eq:3.4}) we suppressed the extra $Q^{2}$ dependence
which at large $Q^{2}$
comes from the scaling violations in the gluon density factor
$\propto G^{2}(x,\tau(Q^{2}+m_{V}^{2}))$, see (\ref{eq:2.8}).
For these scaling violations, at fixed $x_{eff}$ and
asymptotically large $Q^{2}$ we
expect $n\lsim 3$.
In Fig.~7a we present our predictions for $\rho^{0}$ and $\phi^{0}$
production at $W=100\,GeV$ as a function of $Q^{2}+m_{V}^{2}$
assuming for the longitudinal polarization $\epsilon=1$ as in the
ZEUS kinematics \cite{ZEUSrhoQ2}.
These cross sections can be roughly approximated
by the $\propto (Q^{2}+m_{V}^{2})^{-n}$ law with the exponent
$n\approx 2.4$ for the semiperturbative  $r_{S}$ region
$1\lsim Q^{2}\lsim 10$ GeV$^2$. At fixed $W$, $x_{eff}$
varies with $Q^{2}$ and for the $x_{eff}$ dependence of
$\sigma(x_{eff},Q^{2})$ we predict $n\approx 3.2$ for
the perturbative $15\lsim Q^{2}\lsim 100$\,GeV$^{2}$
where $r_{S}$ is small.
We strongly urge a careful analysis of the $Q^{2}$
dependence in terms of the natural variable $Q^{2}+m_{V}^{2}$
(for more discussion see section 5 below).
For the sake of completeness, in Fig.5 we present also
our predictions for the energy dependence of the
polarization-unseparated production cross section
$\sigma=\sigma_{T}+\epsilon\sigma_{L}$ for the typical
$\epsilon=1$.

%-----------------------------------------

%   Section 4

%------------------

\section{Anomalies in electroproduction of $2S$ radially
excited vector mesons}

Here the keyword is the node effect - the
$Q^{2}$ and energy dependent cancellations
from the soft (large size) and hard (small size) contributions
to the production amplitude of the $V'(2S)$ radially excited vector
mesons. When the value of the scanning radius $r_{S}$ is
close to the node $r_{n}\sim R_{V}$,
these cancellations must exhibit a strong
dependence on both $Q^{2}$ and energy due to the
different energy dependence of the dipole cross section at
small ($r<R_{V}$) and large ($r>R_{V}$) dipole sizes.
It must be made clear from the
very beginning that when strong cancellations of the large and
small region contributions are involved, the predictive power becomes
very weak and the results strongly model dependent.
Our predictions for the production of the $V'(2S)$ radial excitations
which we report here
serve mostly as an illustration of the unusual $Q^{2}$
and energy dependence possible in these reactions.
(Manifestations of the node
effect in electroproduction on nuclei were discussed earlier, see
\cite{NNZanom} and \cite{BZNFphi})

In the nonrelativistic limit of heavy quarkonia, the
node effect will not depend on the polarization of the virtual photon
and of the produced vector meson. Not so for light
vector mesons.
The wave functions of the transversely and longitudinally
polarized photons are different, the regions of $z$ which
contribute to the ${\cal M}_{T}$ and ${\cal M}_{L}$ are different, and the
$Q^{2}$ and energy dependence of the node effect in production
of the transverse and longitudinally polarized $V'(2S)$ vector mesons
will be different.

Let us start with the transverse amplitude. Two cases can occur
\cite{NNZanom}, the undercompensation
and the overcompensation scenario.
In the undercompensation case,
the production amplitude
$\langle 2S|\sigma(x_{eff},r)|\gamma^*\rangle$
is dominated by the positive contribution coming from
$r\lsim r_{n}$
and the $V(1S)$ and $V'(2S)$ photoproduction
amplitudes have the same sign. With
our model wave functions this scenario is realized for
transversely polarized
$\rho'$ and $\phi'$
 (we can not, however, exclude the overcompensation scenario).
As discussed in
\cite{NNZanom}, in the undercompensation scenario a decrease of
of the scanning radius
with $Q^{2}$ leads to
a rapid decrease of the negative contribution coming from
large $r\gsim r_{n}$ and to a rapid
rise of the
$V'(2S)/V(1S)$ production ratio with $Q^{2}$.
The stronger the suppression of the real photoproduction of
the $V'(2S)$ state, the steeper the $Q^{2}$ dependence
of the $V'(2S)/V(1S)$ production ratio expected at small $Q^{2}$.
With our model wave
functions, the $\rho'(2S)/\rho^{0}$ and $\phi'(2S)/\phi^{0}$
production ratios for the transverse
polarization are predicted to rise by more than one order of magnitude
in the range $Q^{2}\lsim 0.5$\,GeV$^{2}$, see Fig.~8;
the $V(2S)$ and $V(1S)$
production cross sections become comparable
at $Q^{2}\gsim 1$\,GeV$^{2}$, when the production
amplitudes are dominated by dipole size $r\ll r_{n}$
\cite{NNZanom,NNZscan}.

For the longitudinally polarized $\rho'(2S)$ and $\phi'(2S)$ mesons,
our model wave functions predict overcompensation;
at $Q^{2}=0\,GeV^{2}$ the amplitude is dominated by
the negative contribution from $r\gsim r_{n}$.
Consequently,
with the increase of $Q^{2}$, i.e. with the decrease of the scanning
radius $r_{S}$, one encounters the {\sl exact} cancellation
of the large and small distance contributions. Our model
wave functions lead to this exact node effect
in the dominant imaginary part of the production amplitude
at some value
$Q_{n}^{2}\sim 0.5$\,GeV$^{2}$ for both the $\rho'_{L}(2S)$ and
$\phi'_{L}(2S)$ production (see Fig.~6).
The value of $Q_{n}^{2}$ is
slightly different for the imaginary and the real part of
the production amplitude
but the real part is typically very small
and this difference will be hard to observe experimentally.
Here we can not insist on the precise value of
$Q_{n}^{2}$ which is subject to the soft-hard cancellations,
our emphasis is on the likely scenario with the exact node
effect at a finite $Q_{n}^{2}$.

We wish to emphasize that only the experiment will be able to
decide between the
overcompensation and undercompensation scenarios.
For instance, let the $\rho^{0}$ and
$\rho'(2S)$ be observed in the $\pi \pi$ photoproduction
channel. The
S\"oding-Pumplin effect of interference
between the direct, non-resonant $\gamma p \rightarrow \pi\pi p$
production and the resonant $\gamma p \rightarrow
\rho^{0}(\rho') p\rightarrow
\pi\pi p$ production amplitudes leads to the skewed
$\rho^{0}$ and $\rho'$ mass spectrum. The asymmetry of
the $\rho^{0}(\rho')$ mass spectrum depends on the sign of the
$\rho^{0}(\rho')$ production amplitudes
(\cite{Soding},
the detailed theory has been worked out in \cite{Pumplin}).
The S\"oding-Pumplin technique has already been applied
to the $\rho'(1600)$ mass region in $\gamma p \rightarrow
\pi^{+}\pi^{-}$ at 20 GeV
studied in the SLAC experiment \cite{Abe}. Their
fit to the $\rho'(1600)$ mass spectrum requires that
the sign of the $\rho'$ production
amplitude be negative relative to that of the $\rho$.
Although the interpretation of this result is not clear
at the moment, because
there are two  $\rho'(1450)$ and $\rho'(1700)$ states
which were not resolved in this experiment, the
S\"oding-Pumplin technique
seems promising.

With the further increase of $Q^{2}$ and decrease of the
scanning radius one enters the above described
undercompensation scenario. Although
the radii of the $s\bar{s}$
and $u\bar{u},d\bar{d}$ vector mesons are different,
the $Q^{2}$ dependence of $\rho'(2S)/\rho^{0}$ and
$\phi'(2S)/\phi^{0}$ production cross section ratios
will exhibit a similar pattern. For both the transverse and longitudinally
polarized photons, these ratios rise steeply with $Q^{2}$
on the scale $Q^{2}\sim 0.5$\,GeV$^{2}$.
At large $Q^{2}$ where the production of
longitudinally polarized mesons dominates,  the
$\rho'(2S)/\rho^{0}$ and $\phi'(2S)/\phi^{0}$ cross section ratios
level off at $\sim 0.3$ (see Fig.~8).
This large-$Q^{2}$
limiting value of the
$\rho'(2S)/\rho^{0}$ and $\phi'(2S)/\phi^{0}$ cross section ratios
depend on the ratio of $V'(2S)$ and $V(1S)$ wave functions
at the origin, which in potential models is subject to
the detailed form of the confining potential
\cite{Potential}. It is interesting that due to the
different node effect for the $T$ and $L$ polarizations,
we find $R_{LT}(2S)\ll R_{LT}(1S)$ , see Fig.~6.

In Fig.7b we present our predictions for the
$Q^{2}$ dependence of the
polarization-unseparated cross section
$\sigma(\gamma^{*}\rightarrow V'(2S))=
\sigma_{T}(\gamma^{*}\rightarrow V'(2S))
+\epsilon\sigma_{L}(\gamma^{*}\rightarrow V'(2S))$
at the HERA energy $W=100\,GeV$ assuming $\epsilon = 1$.
In Fig.~9 we show the $Q^{2}$ dependence of the
polarization-unseparated forward cross section ratios
$d\sigma(\gamma^{*}\rightarrow \rho'(2S))/
d\sigma(\gamma^{*}\rightarrow \rho^{0})$
and
$d\sigma(\gamma^{*}\rightarrow \phi'(2S))
/d\sigma(\gamma^{*}\rightarrow \phi^{0})$ at $W=100$\,GeV.
Due to its smallness, the
anomalous properties of $\sigma_{L}(2S)$ at small $Q^{2}$
are essentially
invisible in the polarization-unseparated $V'(2S)$ production
cross section shown in Figs.~7b,~9 and 10.
In contrast to
$\sigma(\gamma^{*}\rightarrow V(1S))$, which falls
monotonically and steeply
from $Q^{2}=0\,GeV^{2}$ on, the
$\sigma(\gamma^{*}\rightarrow V'(2S))$
shown in Fig.~7b
exhibits a weak rise at small $Q^{2}$.
At $Q^{2}$ large enough that the scanning radius $r_{S}<R_{V}$
and the node effect becomes negligible, we predict very similar
dependence on  $Q^{2}+m_{V}^{2}$
of the $V'(2S)$ and $V(1S)$ production.

Color dipole dynamics uniquely is the source of such a tricky
$Q^{2}$ dependence of the $V'(2S)/V(1S)$ production ratio.
We already mentioned about the experimental confirmation
\cite{NMCPsi',E687Psi'}
of the node effect predicted in $\Psi'$ production
\cite{KZ91}.
Further experimental confirmations of the node
effect, in particular  of the unique overcompensation
scenario which is possible for light vector
mesons, would be extremely interesting.
The available experimental data on real
photoproduction of radially excited
light $V'(2S)$ mesons confirm
$\sigma(\gamma\rightarrow V'(2S))/
\sigma(\gamma \rightarrow V(1S)) \ll 1$,
but are still
of a poor quality and the branching ratios of the $V'$ decays
are not yet established (for the review see \cite{Clegg}
and the Review of Particle Properties \cite{PDT}).
For instance, the FNAL
E401 experiment at $\nu \approx 100$ GeV found
\cite{Philownu}
$\sigma(\gamma \rightarrow \phi'(1700, K^{+}K^{-}))=
8.0\pm 2.7(stat)\pm 1.4(syst)$\,nb to be compared with
$\sigma(\gamma \rightarrow \phi)\approx 0.55$\,$\mu$b
(see Fig.4). In the $\rho$ family, the very
spectroscopy of the $\rho'$
mesons is not yet conclusive \cite{Clegg,PDT}.
There are two $\rho'$ states,
$\rho'(1450)$ and $\rho'(1600)$,
the $2S$ and $D$-wave assignment for these
states is not yet clear.
The first high energy data on the $\rho'(1450)$ and
$\rho'(1700)$ leptoproduction were reported by the
E665 collaboration \cite{E665rop}.
These E665 data refer to the coherent production on Ca
target.
For the $\rho'(1700)$, they exhibit a strong rise of
$R_{21}=\sigma(\rho' \rightarrow 4\pi)/\sigma(\rho \rightarrow 2\pi)$
with $Q^{2}$ by more than one order in magnitude from
$(0.004\pm\,0.004)$ at $Q^{2} = 0.15\,GeV^{2}$ to
$(0.15\pm\,0.07)$ at $Q^{2} = 4.5\,GeV^{2}$.
Such a steep $Q^{2}$ dependence is perfectly consistent
with our expectations for the production of radially
excited 2S light vector mesons.
For the $\rho'(1450)$ there is a weak evidence of a
nonmonotonic $Q^{2}$ dependence:
$R_{21} = (0.035\pm\,0.011)$ at $Q^{2}=0.15\,GeV^{2}$ followed by
decrease down to $R_{21} = (0.012\pm\,0.004)$ at $Q^{2}=0.3\,GeV^{2}$
and then to an increase and leveling off to
$R_{21} = (0.08\pm\,0.04)$ at larger $Q^{2} \geq 2\,GeV^{2}$.
Such a $Q^{2}$ dependence of $R_{21}$ would be natural for a D-wave
state which has a nodeless radial wave
function. If these E665 observations will be confirmed
in higher statistics experiments, then the color dipole
interpretation of the $Q^{2}$ dependence would strongly
suggest the
2S and D-wave state assignments for the $\rho'(1700)$ and
$\rho'(1450)$, respectively. We remind the reader that,
for a quantitative comparison with the predictions of our model
(shown in Fig.~10), the E665 results for $R_{21}$ must
be corrected for the branching ratio $B(\rho' \rightarrow 4\pi)$,
which is still experimentally unknown \cite{PDT}.

The energy dependence of the $\rho'(2S),\phi'(2S)$
real photoproduction is shown in Fig.~10 and has its
own peculiarities. In the color dipole gBFKL dynamics,
the negative contribution
to the $2S$ production amplitude coming
from large size dipoles,
$r\gsim r_{n}$,
has a slower growth with energy than the positive
contribution coming
from the small size dipoles, $r\gsim r_{n}$. For this reason,
in the undercompensation regime the destructive interference
of the two contributions becomes weaker at higher energy and
we predict a growth of the
$V'(2S)/V(1S)$ cross section ratios with energy.
Taking only the pure pomeron contributions into account,
we find for the forward cross
section ratio $d\sigma(\gamma\rightarrow \rho'(2S)/
d\sigma(\gamma\rightarrow \rho^{0}) = 0.041$
at $W = 15$~GeV, which at HERA energies increases to 0.063 and
0.071 at $W=100$\,GeV and $W=150$\,GeV, respectively.
Whereas in $\rho$ and $\rho'$ production one must be aware of
the non-vacuum Reggeon exchange
contributions at lower energy, in the pomeron dominated
$\phi',\phi$ real photoproduction we find a somewhat
faster rise of
$d\sigma(\gamma\rightarrow \phi'(2S)/
d\sigma(\gamma\rightarrow \phi^{0})$ with energy from $0.054$
at $W = 15$~GeV to 0.089 and
0.099 at $W=100$\,GeV and $W=150$\,GeV, respectively.

If the leptoproduction of the longitudinally polarized $V_{L}'(2S)$ will
be separated experimentally, we will have a chance of
studying the $Q^{2}$ and energy dependence in the overcompensation
scenario. Start with the moderate energy and
consider $Q^{2}$ very close to $Q_{n}^{2}$ but
still $\lsim Q_{n}^{2}$. In this case the negative
contribution from $r\gsim r_{n}$ takes over in the $V_{L}'(2S)$
production amplitude. With increasing energy, the
positive contribution to the production amplitude
rises faster and ultimately takes over. At some
intermediate energy, there will be an exact cancellation
of the two contributions to the production
amplitude and the longitudinal
$V_{L}'(2S)$ production cross section shall
exhibit a minimum at this energy (the minimum will partly
be filled
because cancellations in the real and imaginary part of the
production amplitude are not simultaneous).
With our model wave functions,
we find such a nonmonotonic energy dependence of the
$\rho_{L}'(2S)$ and $\phi_{L}'(2S)$
production at $Q^{2}\approx 0.5$\,GeV$^{2}$,
which is shown in Figs.~8 and 10. At higher $Q^{2}$ and smaller
scanning radii $r_{S}$ the energy dependence of $V_{L}'(2S)/V_{L}(1S)$
production ratio becomes very weak.

Finally, a brief comment on the $t$-dependence of the
differential cross sections is in order. For
the 1S vector mesons we expect the conventional
diffractive peak with smooth and gentle energy dependence.
For the radially excited vector mesons the $t$-dependence
can be anomalous. The point is that the large size
contribution to the $V'(2S)$ meson production amplitude
has steeper $t$-dependence that the small size contribution.
The destructive interference of these two amplitudes can
lead to two effects:
i) the diffraction slope in the $V'(2S)$
meson production will be smaller than in the $V(1S)$ meson
production,
ii) the effective diffraction slope for
the $V'(2S)$ meson production decreases towards small $t$
contrary to the familiar increase for the $V(1S)$ meson
production. High statistics data on the $\rho',\phi'$
production at HERA are needed to test these
predictions.
More detailed discussion of the diffraction slope
will be presented elsewhere.

%-------------------------------------------

%    Section 5

% -------------

\section{Scaling relations between production of
different vector mesons}

The color dipole cross section is flavor blind and only depends
on the dipole size.
The results (\ref{eq:2.9}),(\ref{eq:2.10}) for the production amplitudes
strongly suggest the restoration of flavor symmetry,
i.e., a similarity between the production of different
vector mesons when compared at the same value of the scanning
radius $r_{S}$
and/or the same value of $Q^{2}+m_{V}^{2}$.
\footnote{For the first considerations of the
restoration of flavor symmetry in diffractive
production of vector mesons off nuclei see \cite{KNNZ94}, the
scaling relations between diffraction slopes for $\gamma^{*}p
\rightarrow Vp$ are discussed in \cite{NZZslope}.}
Such a
comparison must be performed at the same energy, which
also provides the equality of $x_{eff}$ at equal
$Q^{2}+m_{V}^{2}$.
Evidently, the value of $Q^{2}$ must be large enough so that
the scanning radius $r_{S}$ is smaller than the radii of
vector mesons compared.

In order to illustrate the above point we present in Figs. 11 and 12
the ratio of forward production cross sections
$R((J/\Psi)/\rho^{0};Q^{2})=
d\sigma(\gamma^{*}\rightarrow J/\Psi)/
d\sigma(\gamma^{*}\rightarrow \rho^{0})$ and
$R(\phi^{0}/\rho^{0};Q^{2})=
{d\sigma(\gamma^{*}\rightarrow \phi^{0})/
d\sigma(\gamma^{*}\rightarrow \rho^{0})}$
as a function of the c.m.s energy $W$
at different $Q^{2}$
(here we use for the $J/\Psi$ production cross section the values obtained
from a recent calculation \cite{NNPZheavy}, which practically coincide
with those of ref.\cite{NNZscan}, the slight difference being
due to a somewhat different $J/\Psi$ wave function).
Here we compare the polarization-unseparated cross sections
$\sigma=\sigma_{T} + \epsilon \sigma_{L}$, taking
for the definiteness $\epsilon = 1$ which is typical of the HERA
kinematics.
These ratios exhibit quite a strong $Q^{2}$ dependence,
which predominantly comes from the $Q^{2}$ dependence of the
factor
$$
\left({Q^{2}+m_{V_{1}}^{2} \over  Q^{2} + m_{V_{2}}^{2}}\right)^{n}\,,
$$
which changes rapidly when the two vector mesons have different
masses. The  energy dependence of the cross section ratios taken
at the same $Q^{2}$ derives from the different energy dependence
of the dipole cross section which enters at different radii
$r_{Si} \approx \frac{6}{\sqrt{Q^{2}+m_{Vi}^{2}}}$ in the numerator and
denominator of the $V_{1}/V_{2}$ cross section ratio,
$$
R(V_{1}/V_{2};Q^{2})=
{\sigma(\gamma^{*}\rightarrow V_{1})\over
\sigma(\gamma^{*}\rightarrow V_{2})} \propto
{\sigma^{2}(\nu,r_{S1}) \over \sigma^{2}(\nu,r_{S2})} \, .
$$
In the HERA energy range we predict
$R((J/\Psi)/\rho^{0};Q^{2}=0)={\sigma(\gamma\rightarrow J/\Psi)/
\sigma(\gamma\rightarrow \rho^{0})} =0.022$ at $W=70$\,GeV
and 0.028 at $W=150$\,GeV, which agrees with
the experimentally observed ratio $0.0034\pm\,0.0014$
of H1 ($W \sim 70\,GeV$) \cite{H1Psi,H1rho} and
$0.0045\pm\,0.0023$ of ZEUS ($W=150\,GeV$)
\cite{ZEUSrho94,ZEUSrho95,ZEUSPsi}. Notice the rise of
$R((J/\Psi)/\rho^{0};Q^{2})$ by more than 3 orders
in the magnitude from $Q^{2}=0$ to $Q^{2}=100$\,GeV$^{2}$.
Our result for the
ratio $R(\phi^{0}/\rho^{0};Q^{2}=0)=
{d\sigma(\gamma\rightarrow \phi^{0})/
d\sigma_{\Pom}(\gamma\rightarrow \rho^{0})}$ shown in Fig.~11
is substantially smaller than the factor $2/9$ expected from the naive
VDM, in a very good agreement with the experiment
(\cite{Philownu} and references therein).
This suppression is a natural consequence of the
color dipole approach and derives from the smaller
radius of the $s\bar{s}$
quarkonium  and smaller transverse size of
the $s\bar{s}$ Fock state of the photon as compared to
the radius of the $\rho^{0}$ and size of the $u\bar{u},d\bar{d}$ Fock
states of the photon, respectively, cf. Table 1.
For increasing $Q^{2}$s, the
ratio $R(\phi^{0}/\rho^{0};Q^{2})$ overshoots the VDM ratio $2/9$ and
rises by one order of magnitude from
$Q^{2}=0$ to $Q^{2}=100$\,GeV$^{2}$.
In Figs.~11 and 12 we compare only pure pomeron contributions
to the production cross section;
at smaller values of the energy and of the Regge parameter $\omega$, the
$\phi^{0}/\rho^{0}$ and $(J/\Psi)/\rho^{0}$ production ratios will be
further suppressed by the factor $f^{2}$.

The remarkable restoration of flavor symmetry
in the natural scaling variable $Q^{2}+m_{V}^{2}$
is demonstrated in
Figs. 13 and 14, where
we present a ratio $R(i/k;Q^{2}+m_{V}^{2})$
of the same cross sections
taken at equal $Q^{2}+m_{V}^{2}$
rather than equal $Q^{2}$. A marginal variation of the
$R((J/\Psi)/\rho^{0};Q^{2}+m_{V}^{2})$ and
$R(\phi^{0}/\rho^{0};Q^{2}+m_{V}^{2})$ in this scaling
variable must be contrasted with the
variation of the  $R((J/\Psi)/\rho^{0};Q^{2}=0)$ and
$R(\phi^{0}/\rho^{0};Q^{2}=0)$ by the three
and one orders in the magnitude previously mentioned, respectively,
over the same span of $Q^2$ values
$0 < Q^{2}<100$\,GeV$^{2}$.
The origin of the slight departures from exact scaling in the
variable $Q^{2}+m_{V}^{2}$ comes from a well understood difference between
the  scales $A_{T,L}$ and $\tau_{T,L}$ for production
of different vector mesons.
The same difference of $A_{T,L}$ and $\tau_{T,L}$
brings in the energy dependence of $R(i/k;Q^{2}+m_{V}^{2})$. This
is a specific prediction from the preasymptotic gBFKL dynamics.
The radii of the $\phi^{0}$ and $\rho^{0}$ mesons do not
differ much and for this reason we find a precocious
scaling in $Q^{2}+m_{V}^{2}$. The energy dependence
of the $\phi^{0}/\rho^{0}$ ratio also turns out very weak.
The radii of the $\rho^{0}$ and $J/\Psi$ differ
much more strongly and the ratio
$ R((J/\Psi)/\rho^{0};Q^{2}+m_{V}^{2})$ exhibits a somewhat
stronger dependence on energy and
$Q^{2}+m_{V}^{2}$. For the same reason, we
predict a substantial departure of
$R(i/k;Q^{2}+m_{V}^{2})$  from the
short-distance formula
\beq
R(i/k;Q^{2}+m_{V}^{2})={m_{i}\Gamma_{i}(e^{+}e^{-}) \over
m_{k}\Gamma_{k}(e^{+}e^{-}) }\, ,
\label{eq:5.1}
\endeq
which is shown in Fig. 13 by horizontal lines.
The formula (\ref{eq:5.1}) can readily be derived generalizing the
asymptotic-$Q^{2}$ considerations \cite{Ryskin}, for the further
discussion of the crucial r\^ole of the scaling variable
$Q^{2}+m_{V}^{2}$ in this comparison see below Section 6.

The case of the $\omega^{0},\omega'$ virtual photoproduction
is very interesting.
Is the $\rho^{0}-\omega^{0}$ mass degeneracy accidental?
Does it imply also similar spatial wave functions in the
$\rho$ and $\omega$ families?
The scanning property of diffractive production allows a
direct comparison of spatial wave functions of the
$\rho^{0}$ and $\omega^{0}$.
If the $\omega^{0}-\rho^{0}$ degeneracy extends
also to the spatial wave functions, then we predict
\beq
{\sigma(\gamma^{*}\rightarrow \omega^{0})
\over
\sigma(\gamma^{*}\rightarrow \rho^{0}) }
={1\over 9}
\endeq
independent of energy and
$Q^{2}$. On the other hand, if the radii of
the $\rho^{0}$ and $\omega^{0}$ are different, for instance
$R_{\omega}<R_{\rho}$, then
the
$\omega^{0}/\rho^{0}$ production ratio must exhibit the
$Q^{2}$ dependence reminiscent of the $\phi^{0}/\rho^{0}$
ratio.
Similarly, a comparison of the $\omega'$ and $\rho'$ production
can shed light on the isospin dependence of
interquark forces in vector mesons.

%---------------------------------------------------

%     Section 6

%------------------

\section{Determination of color dipole cross section
from vector meson production data}

Inverting
Eqs.~(\ref{eq:2.9}),(\ref{eq:2.10}) one can evaluate
$\sigma(x_{eff},r)$
from the vector meson production data. It is convenient to
cast Eqs.~(\ref{eq:2.9}),(\ref{eq:2.10}) in the form
\beq
{\rm Im}{\cal M}_{T}=g_{T}
\sqrt{4\pi \alpha_{em}}
C_{V}\sigma(x_{eff},r_{S}){m_{v}^{2}\over m_{V}^{2}+Q^{2}}
\label{eq:6.1}
\endeq
\beq
{\rm Im}{\cal M}_{L}=g_{L}
\sqrt{4\pi \alpha_{em}}
C_{V}\sigma(x_{eff},r_{S}){\sqrt{Q^{2}}\over m_{V}} \cdot
{m_{v}^{2}\over m_{V}^{2}+Q^{2}}
\label{eq:6.2}
\endeq
In (\ref{eq:6.1}),(\ref{eq:6.2})
the coefficient functions $g_{T,L}$ are defined so as
to relate the amplitude to $\sigma(r_{S})$ at the
well defined scanning radius (\ref{eq:1.2}) with $A\equiv 6$.
The major point of this decomposition is that at large $Q^{2}$
and/or small $r_{S} \lsim R_{V}$,
the coefficients $g_{T,L}$ will be very smooth functions of
$Q^{2}$ and energy. The smooth $Q^{2}$ and energy dependence
of $g_{T,L}$ mostly reflects the smooth and well understood
$Q^{2}$ dependence of
scale factors $A_{T,L}$.
Such a procedure is somewhat crude and the
${\rm Im}{\cal M}_{T,L}-\sigma(x_{eff},r_{S})$
relationship is sensitive to the assumed
$r$ dependence of the dipole cross section $\sigma(x_{eff},r)$.
Using the dipole cross section \cite{NNZscan}
the shape of which changes significantly
from $\omega = {1\over x_{eff}}=30$ up to
$\omega = 3\cdot 10^{6}$,
we have checked that this
sensitivity is weak.
In Fig.~15 we
present the $Q^{2}$ dependence of the $g_{T,L}$ for different
production processes at $W=15$\,GeV and $W=150$\,GeV.
The variation of the
resulting coefficient functions $g_{T,L}$ from small to
large $W$ does not exceed $15\%$, which is
a conservative estimate of the theoretical
uncertainty of the above procedure.

The experimentally measured forward cross production
section section equals
\beq
\frac{d\sigma(\gamma^{*} \rightarrow V)}{dt}|_{t=0} =
\frac{f^{2}}{16\pi}\cdot\Biggl[(1+\alpha_{V,T}^{2}){\cal M}_{T}^{2} +
\epsilon(1+\alpha_{V,L}^{2}){\cal M}_{L}^{2}\Biggr]
\label{eq:6.3}
\endeq
The difference between $\alpha_{V,L}$
and $\alpha_{V,T}$ for the longitudinal
and transverse cross sections
and the overall effect of the real part
is marginal and can safely
be neglected compared to other uncertainties.
Then,
making use of the above determined $g_{T,L}$ and combining
Eqs.~(\ref{eq:6.1}),
(\ref{eq:6.2}) and (\ref{eq:6.3}), we obtain
\arr
\sigma(x_{eff},r_{S}) = \frac{1}{f}\cdot\frac{1}{C_{V}}\cdot
{Q^{2}+m_{V}^{2} \over m_{V}^{2}}
\cdot
{2 \over \sqrt{\alpha_{em}}}\cdot
\left(
g_{T}^2 +\epsilon\,g_{L}^{2}\cdot{Q^{2}\over M_{V}^{2}}\right)^{-1/2}
\cdot     \nonumber \\
\cdot
\left(
1 +\alpha_{V}^{2}\right)^{-1/2}
\sqrt{\left.{d\sigma(\gamma^{*}\rightarrow V) \over dt}
\right|_{t=0}}
\label{eq:6.4}
\endarr
Here $\epsilon$ is the longitudinal polarization of the photon
the values of which are taken from the corresponding experimental
publications.
In (\ref{eq:6.3}),(\ref{eq:6.4}) $f$ is the above discussed factor
which accounts for the non-vacuum Reggeon contribution to
the $\rho^{0}$ production, for $\phi^{0}$ and $J/\Psi$
production, $f \equiv 1$.
In the case the experimental data are presented in the form of
the $t$-integrated cross section, we evaluate
$
\left.{d\sigma(\gamma^{*}\rightarrow V) \over dt}
\right|_{t=0}   = B\sigma_{tot}(\gamma^{*}\rightarrow V)
$ using the diffraction slope $B$ as cited in
the same experimental publication.

In Fig.~16  we show the results of
such an analysis on the low energy \cite{Philownu}
and ZEUS \cite{ZEUSphi}
$\phi^{0}$ real photoproduction data,
on the $\rho^{0}$ and
$\phi^{0}$ NMC electroproduction data \cite{NMCfirho},
on the $\rho^{0}$ HERA real and
virtual photoproduction (H1 \cite{H1rho},
ZEUS \cite{ZEUSrho94,ZEUSrho95,ZEUSrhoQ2}), on the fixed target data on
real photoproduction (EMC \cite{EMCPsi}, E687 \cite{E687}),
on the EMC $J/\Psi$ electroproduction data
(\cite{EMCPsiQ2}) and on the HERA
real photoproduction $J/\Psi$ data
(H1 \cite{H1Psi}, ZEUS \cite{ZEUSPsi}).
The error bars are the error bars in the
measured cross sections as cited in the experimental
publications. The experimental
data on the vector meson production give a solid evidence
for a decrease of
$\sigma(x_{eff},r_{S})$ by one order of magnitude from $r_{S}
\approx 1.2$ fm in $\phi^{0}$ real photoproduction
down to $r_{S}\approx 0.24$ fm in the  electroproduction of
$\rho^{0}$ at $Q^{2}=23$\,GeV$^{2}$ and of $J/\Psi$
at $Q^{2}=13$\,GeV$^{2}$. In the region of overlapping
values of $r_{S}$ there is a
remarkable consistency between
the dipole size dependence and the absolute
values of the
dipole cross section determined from the data on the
$\rho^{0},\phi^{0}$ and $J/\Psi$ production, in agreement with
the flavor independence of the dipole cross section.
A comparison of determinations of $\sigma(x_{eff},r)$ at
fixed-target and
HERA energy
confirms the prediction \cite{NZZ94,NZBFKL,NNZscan} of
faster growth of the dipole cross section at smaller dipole
size, although the error bars are still large.

The above determination of $\sigma(x_{eff},r_{S})$ is rather
crude for the several reasons. \\
i)
First, a comparison of the
NMC \cite{NMCfirho} and early EMC data \cite{EMCrho} on
the $\rho^{0}$ production suggests that the admixture
of inelastic process
$\gamma^{*}p \rightarrow VX$
could have enhanced the EMC cross section by as large a
factor as $\sim 3$ at $Q^{2} = 17\,GeV^{2}$.
The value of $\sigma(\gamma^{*}\rightarrow V)$ thus
overestimated, leads to
$\sigma(x_{eff},r_{S})$ overestimated by the factor
$\sim \sqrt{3}$, which
may be a reason why
the EMC $J/\Psi$ electroproduction data
\cite{EMCPsiQ2}
lead consistently to somewhat larger values of
$\sigma(x_{eff},r_{S})$.
Still, even this factor of $\sim\sqrt{3}$
 uncertainty is much smaller
than the more than the one order of magnitude by which
$\sigma(x_{eff},r_{S})$ varies over the considered span of $r_{S}$.
In the recent NMC data \cite{NMCfirho} a special care has
been taken to eliminate an inelastic background and the values
of $\sigma(x_{eff},r_{S})$ from the $\rho^{0}$ and
$\phi^{0}$ production data are consistent within the experimental
error bars. \\
ii)
There are further uncertainties with the value of the
diffraction slope $B(\gamma^{*}\rightarrow V)$ and
the curvature of the diffraction cone which affect the
extrapolation down to $t = 0$. The experimental
situation with the diffraction slopes is quite unsatisfactory;
in the case of the $J/\Psi$ and of the light vector mesons
at large $Q^{2}$, one can not exclude even a $\sim 50\%$ uncertainty
in the value of  $B(\gamma^{*}\rightarrow V)$.
However, this
uncertainty in $B(\gamma^{*}\rightarrow V)$
corresponds to
$\lsim 25\%$ uncertainty in our evaluation of
$\sigma(x_{eff},r_{S})$, which is sufficient
for the purposes of the present exploratory study. \\
iii)
In addition, there is also
the above evaluated
conservative $\lsim$ 15\% theoretical inaccuracy of our
procedure.  \\
iv)
Finally, there is a residual uncertainty concerning the wave function
of light vector mesons. As a matter of fact, if the
dipole cross section were known, the diffractive
production $\gamma^{*}p\rightarrow Vp$ would be a
unique local probe of the vector meson wave function
at $r\approx r_{S}$ \cite{KNNZ93}; this may well become
one of the major applications of vector meson production.
To this aim, the consistency of $\sigma(x_{eff},r)$
determined from different reactions indicates that wave
functions of vector mesons are reasonably constrained by
modern spectroscopic models and by
the leptonic width.

This is the first direct determination of the dipole cross
section from the experimental data and our main conclusions
on the properties of the dipole cross section are not affected
by the above cited uncertainties.
In Fig.~16 we show also the dipole cross section from the
gBFKL analysis \cite{NZHera,NNZscan}, which gives a good
quantitative description of structure function
of the photon
at small $x$. We conclude that the color dipole
gBFKL dynamics provides a unified description of
diffractive production of vector
mesons and of the proton structure function.

%kolya begin
{%\bf
Finally, a  comparison of the color dipole analysis of
diffractive electroproduction \cite{KNNZ93,KNNZ94,NNZscan}
with the related momentum space analysis
of Refs.~\cite{Ryskin,Brodsky} is in order.
At a very large $Q^{2}$ and/or very short scanning radius,
$r_{S} \ll R_{V}$,
the electroproduction probes the wave function of vector mesons
and/or the z-distribution amplitude
at a vanishing transverse size, integrated over the
$z$ with the certain $z$-dependent factor which emerges in
Eqs.~(\ref{eq:2.4}),(\ref{eq:2.5}). The wave function at the
vanishing 3-dimensional separation of the quark and
antiquark can be
related to the width of the leptonic decay, $V\rightarrow
e^{+}e^{-}$. The form of the $z$-dependent factor is mostly
dictated by  $r_{S}^{2} \sim 1/(4z(1-z)Q^{2}+m_{V}^{2})$
which emerges in the integrands of (\ref{eq:2.4}),(\ref{eq:2.5})
after the $r$ integration, and for the asymptotical $Q^{2}$ when
$4z(1-z)Q^{2} \gg m_{V}^{2}$,
Brodsky et al. \cite{Brodsky} introduced the moment of
the longitudinal
distribution amplitude
\beq
\eta_{V} ={\int_{0}^{1}dz {1\over 2z(1-z)}\Psi_{V}(r=0,z)
\over
\int_{0}^{1}dz \Psi_{V}(r=0,z)}\,.
\label{eq:6.5}
\endeq
One must be careful with the interpretation of $\eta_{V}$
, though, because for the very asymmetric $q\bar{q}$
configurations, $z(1-z) \lsim M_{V}^{2}/Q^{2}$, the scanning radius
stays large even for $Q^{2}\rightarrow \infty$;
for instance, precisely these
asymmetric configurations dominate the cross section of the diffraction
dissociation of photons, $\gamma^{*} p \rightarrow Xp$, into the
continuum states $X$ \cite{NZ91,NZ94}. With these reservations, we can
combine the representations (\ref{eq:6.1}),(\ref{eq:6.2}),
the pQCD relationship (\ref{eq:2.8}) and the formula
(\ref{eq:1.2}) for the scanning radius,
and cast the production amplitude in the form (here we
focus on the dominant longitudinal amplitude)
\arr
{\cal M}_{L} = {8\pi^{2} \over 3}
 f_{V} \sqrt{4\pi\alpha_{em}}
 \eta_{V}m_{V}
        {\sqrt{Q^{2}} \over m_{V}}
        {1 \over(Q^{2}+m_{V}^{2})^{2}}\alpha_{S}(\bar{Q}^{2})G(x,\bar{Q}^{2})
\nonumber\\
 = {2f_{V} e \eta_{V}m_{V} \over 9}
        {\sqrt{Q^{2}} \over m_{V}}\left({6 \over A}\right)^{2}
  r_{S}^{2} \sigma(r_{S})
\, ,
\label{eq:6.6}
\endarr
where
\beq
f_{V}^{2} = {3\over 8\pi \alpha_{em}^{2}}
            \Gamma(V\rightarrow e^{+}e^{-}) m_{V}\, .
\label{eq:6.7}
\endeq
Then, we can present our results for ${\cal M}_{L}$ in terms of
this parameter $\eta_{V}$. The first line of Eq.~(\ref{eq:6.6})
gives the asymptotic-$Q^{2}$ form of ${\cal M}_{L}$ in terms of
the gluon structure function of the proton, the second line
is equivalent to it at large pQCD factorization scale $\bar{Q}^{2}$
and serves as a working definition of $\eta_{V}$ at moderately
large $Q^{2}$ and/or moderately small scanning radius $r_{S}$.
The above finding that $g_{T,L}$ only weakly depend on $Q^{2}$
and energy, already suggests the $\eta_{V}$ defined by the second
line of Eq.~(\ref{eq:6.6}) will be approximately constant, and
now we show this is indeed the case.

Evidently, the resulting values of
$\eta_{V}$ will depend on the pQCD factorization scale $\bar{Q}^{2}$.
The scale parameter $\tau$ in the pQCD factorization scale
$\bar{Q}^{2}=\tau (Q^{2}+m_{V}^{2})$ was evaluated in \cite{NNZscan}.
It is related to the
scale parameter $B\approx 10$ in the pQCD formula (\ref{eq:2.8})
and the scale parameter $A$ in the scanning radius (\ref{eq:1.2}) as
$\tau \sim B/A^{2}$, for production of the longitudinal
vector mesons in DIS Ref.~\cite{NNZscan} finds
$\tau_{L}(J/\Psi) \approx 0.2$ and $\tau_{L}(\rho^{0})\approx 0.15$.
Ref.~\cite{Brodsky} cites the asymptotic leading twist
form of (\ref{eq:6.6}), with $m_{V}^{2}$ neglected in the
denominator compared to $Q^{2}$ and with the pQCD factorization scale
$\bar{Q}^{2}=Q^{2}$, besides the more accurate definition of the
pQCD scale $\bar{Q}^{2}$
we differ from Ref.~\cite{Brodsky} also
by the factor 2 in Eq.~(\ref{eq:2.8}). The scale $A$ in the scanning
radius is given by the position of the peak in $W_{L}(Q^{2},r^{2})$, it
varies with $Q^{2}$ slightly bringing the slight variation of
the scale factors $A_{T,L}$ in, at large $Q^{2}$ it is reasonable
to take $A_{L}(J/\Psi)=6$ and $A_{L}(\rho^{0},\phi^{0})=8$ \cite{NNZscan}.
With this choice of $A_{L}$, our results for the production amplitude
${\cal M}_{L}$ correspond to
the values of $\eta_{V}$ shown in Figs.~17 and 18.
For the nonrelativistic quarkonium, in which $z\approx {1\over 2}$,
Eq.~(\ref{eq:6.5}) gives $\eta_{V}\approx 2$. The $\Upsilon$ is
a good approximation to the nonrelativistic quarkonium and we
indeed find $\eta_{V}\approx 2$. Taking a fixed scale $A$, we
neglected the slight variation of $\tau$ with $Q^{2}$,
which propagates into the slight variation of $\eta_{V}$
with $Q^{2}$. Because the shape of the color
dipole cross section varies with $x_{eff}$, the scale parameter $A$
varies also with $x$ slightly. Taking the $x$-independent
$A$, we cause the slight mismatch of the $x$-dependence of
the r.h.s. and l.h.s. of Eq.~(\ref{eq:6.5}), which
propagates into the weak $x_{eff}$ dependence of $\eta_{V}$.
In Fig.~17 we show $\eta_{V}$ for the fixed energy $W=150$\,GeV relevant
to the HERA experiments, here the combined $Q^{2}$ and $x_{eff}$
dependence of the scale parameter $A$ contribute to the variations
of $\eta_{V}$. The issue of the $Q^{2}$ and $x$ dependence of the
pQCD factorization scale $\bar{Q}^{2}$
in (\ref{eq:6.5}) deserves a dedicated
analysis, here we only wish to focus on the fact that
the so determined $\eta_{V}$ exhibits a remarkably weak
variation with $Q^{2}$ and $x_{eff}$. Furthermore, Fig.~18 shows
that the $x_{eff}$-dependence
of $\eta_{V}$ becomes substantially weaker at larger $Q^{2}$.
This testifies to an importance of the $Q^{2}+m_{V}^{2}$ as a relevant
scaling variable, which absorbs major mass corrections to the
$Q^{2}$ dependence of the production amplitude (see also the
discussion of the flavor symmetry restoration in Section 5).
To this end we wish to notice that the expansion
\beq
{1\over (Q^{2}+m_{V}^{2})^{2}} = {1\over Q^{4}}
\left( 1+ {2m_{V}^{2} \over Q^{2}}-...\right)
\label{eq:6.8}
\endeq
corresponds to the abnormally large scale $2m_{V}^{2}$ for the
higher twist correction to the production amplitude of leading twist.
For the light vector mesons, Brodsky et al. cite estimates
$\eta_{V}=$3-5, our results in Figs.~17 and 18 are very close to
these estimates, as it must be expected because the momentum-space
technique of Brodsky et al. and our color dipole factorization
technique are related by the Fourier-Bessel transform.
With the present poor knowledge of the
large dipole distributions in vector mesons and/or the wave functions
of vector mesons, the variations of $\eta_{V}$ in Figs.~17 and 18
and the range of estimates for $\eta_{V}$ in \cite{Brodsky}
indicate the range of uncertainty in predictions leptoproduction
amplitudes.
}
%kolya end

%---------------------------------------------

% Section 7

% -------------
\section{ Conclusions}

The purpose of this paper has been the phenomenology of diffractive
photoproduction and electroproduction of ground state ($1S$) and
radially excited ($2S$) light vector mesons in the framework of the
color dipole picture of the QCD pomeron.
In this picture, the
$Q^{2}$ dependence of production of the $1S$ vector mesons is
controlled by the shrinkage of the transverse size of the virtual
photon and the small dipole size dependence of the color dipole cross
section.
Taking the same color dipole cross section as used in the
previous successful prediction of the low $x$ structure function
of the proton, we have obtained a good quantitative description of
the experimental data on diffractive photoproduction and
electroproduction of $1S$ vector mesons $\rho^{0},\phi^{0}$
and $J/\Psi$.
We have presented the first determination of the dipole
cross section from these data and found a remarkable consistency
between the absolute value and the dipole size and energy
dependence of the dipole cross section extracted from the data on
different vector mesons.
This represents an important cross-check
of the color dipole picture.
The pattern we found for the energy dependence
of the dipole cross section is consistent
with flavor independence and
with expectations from the gBFKL dynamics.
The color dipole
picture leads to the restoration of the flavor symmetry and to
novel scaling relations between the production
of different vector mesons when compared at the same
$Q^{2}+m_{V}^{2}$. Such relations are borne out by the available data
and will be further tested when the higher precision data from
HERA will become available.
Regarding this $(Q^{2}+m_{V}^{2})$-scaling, perhaps still
more interesting are the deviations from scaling, which
originate from a substantial contribution of the large size dipoles
even at very large $Q^{2}$s.

The second class of predictions concerns the
rich pattern of an anomalous $Q^{2}$ and energy dependence
of the production of
the $V'(2S)$ radially excited vector mesons,
which depends entirely on the quantum mechanical fact that
the $2S$ wave function has a node which makes these anomalies
an unavoidable effect.
We find a very strong suppression of the
$V'(2S)/V(1S)$ production ratio in the real photoproduction limit of
very small $Q^{2}$.
For the longitudinally polarized $2S$
mesons we find a plausible overcompensation scenario leading to a sharp
dip of the longitudinal cross section
$\sigma_{L}(2S)$ at some finite $Q^{2}
=Q_{n}^{2}\sim 0.5$\,GeV$^{2}$. The position $Q_{n}^{2}$ of this dip
depends on the energy and leads to a nonmonotonic energy
dependence of $\sigma_{L}(2S)$ at fixed $Q^{2}$.
Regarding the experimental choice between the overcompensation and
undercompensation scenarios in the HERA experiments, the
situation looks quite favorable because
the sign of the $\rho'$ production amplitude relative to
that of the $\rho^{0}$ can be measured directly by the
S\"oding-Pumplin method.
At larger $Q^{2}$, the scanning radius becomes shorter,
and we predict a steep rise of the $2S/1S$ cross section ratio,
typically by one order of magnitude
on the very short
scale $Q^{2}\lsim 0.5\,GeV^{2}$
in agreement with the present indications from the E665 data.
The flattening of this $2S/1S$ ratio at large $Q^{2}$ is
a non-negotiable prediction from the color dipole dynamics.
Remarkably, the $Q^{2}$ dependence of the $V'$ production
offers a unique possibility of distinguishing between
$2S$ radially excited and $D$-wave vector mesons.

Finally, in the color dipole framework,
a comparison of the $Q^{2}$ dependence of the diffractive
production of the
$\rho^{0}$ and $\omega^{0}$ constitutes a direct comparison
of the spatial wave functions of the two mesons. A comparison
of the $Q^{2}$ dependence of the $\omega'$ and $\rho'$
production can shed light on the isospin dependence of
the interquark forces in vector mesons.
\medskip\\
{\bf Acknowledgments:} The stay of B.G.Z. at Institut f. Kernphysik,
KFA J\"ulich was supported by DFG. The support by the INTAS
grand 93-238 is acknowledged.
\pagebreak

%=============================

{\bf \large Appendix.}
\bigskip\\

Here we present the parameterization of the wave
functions of vector mesons in the lightcone mixed $({\bf r},z)$
representation.
Due to the fact that small size $q\bar q$ configurations
become important at large $Q^{2}$, one needs
to include the short distance hard
QCD gluon exchange effects so as to make
the electromagnetic form factors consistent with the
QCD predictions.
%kolya begin
{%\bf
Here we follow a simple procedure suggested in \cite{NNZscan},
which uses the relativization technique of
Refs. \cite{Terentiev,BrodskyFF}. We are perfectly aware
of the fact the wave functions of light vector mesons are
still unknown; in the present exploratory
study our major concern is to have a parameterization which
is consistent with the size of vector mesons as suggested by
the conventional spectroscopic models and has the short
distance behavior driven by the hard QCD gluon exchange
\cite{BrodskyFF}.
}
%kolya end

Let the $Vq\bar{q}$ vertex be $\Gamma\bar{q}V_{\mu}\gamma_{\mu}q$
where the vertex function $\Gamma$ is a function of the
lightcone invariant variable \cite{Terentiev}
%------------------------------------------------
\beq
{\bf{p^2}} = \frac{1}{4}(M^{2}-4m_q^2)  \,
\label{eq:A.1}
\endeq
%------------------------------------------------
where
$M$ is the invariant mass of the $q\bar q$ system
%------------------------------------------------
\beq
M^{2} = {m_{q}^{2}+{\bf k}^{2} \over z(1-z)}\, ,
\label{eq:A.2}
\endeq
%------------------------------------------------
${\bf k}$ and $m_{q}$ are the transverse momentum and quark mass,
and $z$ is a fraction of lightcone momentum of the meson carried by
the quark ($0 < z < 1$). In the nonrelativistic limit $\bf p$ is
the 3-momentum of the quark and we have the familiar relationship
between the vertex function and the momentum space wave function
%------------------------------------------------
\beq
\Psi({\bf{p}}^{2})\propto {\Gamma({\bf{p}}^{2})\over
4{\bf{p}}^{2}+4m_{q}^{2}-m_{V}^{2}} \, .
\label{eq:A.3}
\endeq
%------------------------------------------------

The hard gluon exchange Coulomb
interaction ${4\over 3}{\alpha_{S}(d)\over d}$, where
${\bf d}$
is the 3-dimensional quark-antiquark separation and
$\alpha_{S}(d)$ is the running QCD coupling in the
coordinate representation, is singular at the origin,
${\bf d} \rightarrow 0$, but becomes important only at
short distances $d$, much smaller than the radius
$R_{V}$ of the vector meson.
For this reason, the hard gluon Coulomb interaction can be treated
perturbatively.
Namely, let $\Psi_{soft}({\bf d})$ be
the wave function of the vector meson in the soft,
non-singular potential. Solving the Schr\"odinger
equation at small ${\bf d}$ to the first order in Coulomb
interaction, one readily finds the Coulomb-corrected
wave function of the form
%------------------------------------------------
\beq
\Psi({\bf d}) = \Psi_{soft}({\bf d}) +
\Psi_{soft}(0)C\exp \biggl ({ - \frac{d}{2Ca(d)}} \biggr )\,.
\label{eq:A.4}
\endeq
%------------------------------------------------
Here
$a(d)$ is the "running Bohr
radius" equal to
%------------------------------------------------
\beq
a(d) = {3 \over 8 m \alpha_{s}(d)}
\label{eq:A.5}
\endeq
%------------------------------------------------
where $m = m_q/2$ is the reduced quark mass. The parameter
$C$ is controlled by the transition between the hard Coulomb
and the soft confining interaction; we treat it as a variational
parameter. (Similar analysis of the correction to the
momentum space wave function
for the short distance Coulomb interaction is reviewed
in \cite{BrodskyFF}).
The 3-dimensional Fourier transform of the Coulomb-corrected wave
function (\ref{eq:A.4}) reads
%------------------------------------------------
\arr
\Psi({\bf p}) =
N_{0} \Biggl\{ (2\pi R^{2})^{3/2} \exp \biggl [ -{1
\over
2} p^{2}R^{2} \biggr ]
+ C^{4}~ {64
{a^{3}(p^2)}\pi \over ( 1+4{C^{2}a^{2}(p^2)}p^{2})^{2}} \Biggr \}  ,
\label{eq:A.6}
\endarr
%------------------------------------------------
where $a(p^2)$ is still given by (\ref{eq:A.5}) with the running
$\alpha_{S}(p^2)$ evaluated in the momentum representation.

The relativistic lightcone wave function $\Psi(z,\bf{k})$ is obtained
from $\Psi(\bf{p})$ by the standard substitution of the light
cone expression (\ref{eq:A.1},\ref{eq:A.2}) for the nonrelativistic
$\bf{p}^{2}$ in (\ref{eq:A.6})
\cite{Terentiev,BrodskyFF}.
The relativistic wave function
thus obtained gives the correct QCD asymptotics $\propto
\alpha_{S}(Q^{2})/Q^{2}$
of the vector meson form factor, in perfect
correspondence to the familiar hard QCD mechanism (for the
review see \cite{BrodskyFF}; the more detailed analysis of form factors will be
presented elsewhere). Then, the lightcone radial wave function
is the Fourier transform
%------------------------------------------------
\beq
\phi(r,z)=\int {d^{2}{\bf{k}} \over (2\pi)^{2}}\Psi(z,{\bf{k}})
\exp(i{\bf{k}}{\bf{r}}) \,.
\label{eq:A.7}
\endeq
%------------------------------------------------
With the conventional harmonic oscillator form of
$\Psi_{soft}({\bf d })$ we obtain the simple
analytical formula
%------------------------------------------------
\arr
\phi_{1S}(r,z) &=& \Psi_{0}(1S)\Biggl \{ 4z(1-z) \sqrt {2\pi R_{1S}^{2}}
\exp\left[-{m_{q}^{2}R_{1S}^{2} \over 8z(1-z)}\right]
\exp\left[-{2z(1-z)r^{2} \over R_{1S}^{2}}\right]
\exp\left[{m_{q}^{2}R_{1S}^{2} \over 2}\right] \nonumber \\
& + &C^{4}~ {16a^{3}(r) \over AB^{3}} r K_{1}(\beta r)
\Biggr \}
\label{eq:A.8}
\endarr
%------------------------------------------------
where $a(r)$ is given by Eq.~(\ref{eq:A.5}), $\beta = A/B$, and
%------------------------------------------------
\beq
A^{2} = 1 + {C^{2}a^{2}(r)m_{q}^{2} \over z(1-z)} -
4C^{2}a^{2}(r)m_{q}^{2}
\nonumber
\label{eq:A.9}
\endeq
%------------------------------------------------
%------------------------------------------------
\beq
B^{2} = {C^{2}a^{2}(r) \over z(1-z)}
\nonumber
\label{eq:A.10}
\endeq
%------------------------------------------------
For the $1S$ ground state vector mesons we determine the parameters
$R_{1S}^{2}$ and $C$ by
the standard variational procedure using the conventional
linear+Coulomb potential models \cite{Potential}. We check that
the resulting wave function are consistent with the experimentally
measured width of the $V\rightarrow e^{+}e^{-}$ decay (see Tab.~1).
%kolya begin
{%\bf
This is one of the major constraints because at very large $Q^{2}$
and/or $r_{S} \ll R_{V}$, the electroproduction amplitude is
controlled by the wave function at the vanishing transverse size.
For the
heavy quarkonia, we check that the radii of the $1S$ states are
close to the results of more sophisticated solution of the
Schr\"odinger equation \cite{Potential}. The radius of the $\rho^{0}$
 meson given
by our wave function is consistent with the charge radius of the
pion. Still another cross check is provided by
$\sigma_{tot}(\rho^{0}N)$ discussed in Section 3, which
comes out very close to the pion-nucleon total cross section.
}
%kolya end

The node of the radial wave function of the $V'(2S)$
is expected at $r_{n}\sim R_{V}$ far beyond the
Coulomb region. For this reason, we only modify the soft
component of the wave function and take the same
functional form of the
Coulomb correction as for the $1S$ state:
%------------------------------------------------
\arr
\phi_{2S}(r,z) &=& \Psi_{0}(2S)\Biggl \{ 4z(1-z) \sqrt {2\pi R_{2S}^{2}}
\exp\left[-{m_{q}^{2}R_{2S}^{2} \over 8z(1-z)}\right]
\exp\left[-{2z(1-z)r^{2} \over R_{2S}^{2}}\right]
\exp\left[{m_{q}^{2}R_{2S}^{2} \over 2}\right] \nonumber \\
& & \Bigl \{1-\alpha\Bigl [1 + m_{q}^{2}R_{2S}^{2} -
{m_{q}^{2}R_{2S}^{2} \over 4z(1-z)}
+ {4z(1-z) \over R_{2S}^{2}}r^{2} \Bigr ] \Bigr \} \nonumber \\
& + &C^{4}~ {16a^{3}(r) \over AB^{3}} r K_{1}\beta r)
\Biggr \} \Biggr \}\,.
\label{eq:A.11}
\endarr
%------------------------------------------------
The new parameter $\alpha$ controls the position $r_{n}$ of the node.
The two parameters $\alpha$ and $R_{2S}$ are determined from the
orthogonality condition
%----------------------------------------------------------------------
\arr
{N_{c} \over 2\pi}\int_{0}^{1}{ dz \over z^{2}(1-z)^{2} }
\int d^{2}{\bf{r}} \cdot ~~~~~~~~~~~~~~~~~~~~~~~~~~\nonumber\\
\cdot\left\{m_{q}^{2} \phi_{i}(r,z)\phi_{k}(r,z)+[z^{2} +(1-z)^{2}]
[\partial_{r}\phi_{i}(r,z)]
[\partial_{r}\phi_{k}(r,z)] \right\} & = & \delta_{ik}
\label{eq:A.12}
\endarr
%----------------------------------------------------------------------
and from the $2S-1S$ mass splitting evaluated with the same linear+Coulomb
potential. For the heavy quarkonia, we can check the resulting
$V'(2S)$ wave function against the accurate
data on the width of the $V'(2S)\rightarrow e^{+}e^{-}$ decay, the
agreement in all the cases is good.
The so determined parameters, the quark masses used
and some comparisons with the experiment are summarized in Table 1.
It is  $Ca(r)$ which defines at which radii the interaction is
important. A posteriori, for light vector mesons C is found
small, the radius $Ca(r)$ is indeed small and the
resulting parameters are consistent with the assumption that
the Coulomb interaction is a short-distance perturbation. Furthermore,
for the light vector mesons we find $R_{1S}\approx R_{2S}$.
For heavier mesons $C$ is larger and Coulomb effects are
becoming more important  and $R_{2S}>R_{1S}$ in the ratio
closer to the one for the Coulomb system (see Table 1).
Because our Ansatz for the relativistic
wave function has the correct short-distance QCD behavior and
gives a reasonable description of soft cross sections, we
believe it provides a reasonable interpolation between the
soft and hard regimes in the electroproduction of vector mesons.

\pagebreak

\pagebreak
{\bf Figure captions:}
\begin{itemize}

\item[Fig.~1]
~- The color dipole model predictions for
the total cross section $\sigma_{tot}(VN)$
for the interaction of the light
vector mesons $\rho, \rho', \phi$ and $\phi'$ with the nucleon
target as a function of c.m.s. energy $W$.

\item[Fig.~2]
~- The color dipole model
predictions for the $Q^2$ dependence of the observed
cross section $\sigma(\gamma^{*}\rightarrow V)=
\sigma_{T}(\gamma^{*}\rightarrow V)+\epsilon
\sigma_{L}(\gamma^{*}\rightarrow V)$
of exclusive $\rho^0$ and $\phi^0$
production vs. the low-energy NMC \cite{NMCfirho}
and high-energy  ZEUS
\cite{ZEUSrhoQ2} and H1 \cite{H1rho} data.
The top curve is a prediction for the $\rho^{0}$
production at $W=70$\,GeV, the lower curves are for the
$\rho^{0},\phi^{0}$ production at $W=15$\,GeV.
 The dashed curve for the $\rho^{0}$ shows the pure pomeron contribution
$\sigma_{\Pom}(\gamma^{*}\rightarrow \rho^{0})$, the
solid curve for the $\rho^{0}$
shows the effect of correcting for the non-vacuum
Reggeon exchange as described in the text.

\item[Fig.~3]
~- The color dipole model energy dependence predictions
for forward real photoproduction
of $\rho^{0}$ mesons compared with fixed target
data \cite{Rholownu} and high energy datum from the ZEUS
experiment at HERA collider \cite{ZEUSrho94,ZEUSrho95}.
The dashed curve is the pure pomeron exchange contribution,
the solid curve shows the correction for the the non-vacuum
Reggeon exchange as described in the text.

\item[Fig.~4]
~- The color dipole model
predictions for the  energy dependence of real photoproduction
of the $\phi^{0}$ mesons compared with fixed target
\cite{Philownu} and high energy  ZEUS  data (open square
for the $\phi^{0}$ \cite{ZEUSphi}, solid circle for the
$\rho^{0}$ \cite{ZEUSrho94,ZEUSrho95}).

\item[Fig.~5]
~- The color dipole model
predictions of the forward differential cross sections
$d\sigma_{L,T}(\gamma^{*} \rightarrow V)/dt|_{t=0}$ for
transversely (T)  (top boxes) and longitudinally (L)
(middle boxes) polarized
$\rho^0$ and $\phi^0$ and for the polarization-unseparated
$d\sigma(\gamma^* \rightarrow V)/dt|_{t=0}=
d\sigma_{T}(\gamma^* \rightarrow V)/dt|_{t=0}+\epsilon
d\sigma_{L}(\gamma^* \rightarrow V)/dt|_{t=0}$
(bottom boxes) for
$\epsilon = 1$
as a function of the c.m.s. energy $W$
at different values of $Q^2$.

\item[Fig.~6]
~- The color dipole model
predictions for the $Q^2$ and $W$ dependence of the
ratio of the longitudinal and transverse differential cross
sections in the form of the quantity
$
R_{LT}={m_{V}^{2} \over Q^{2}}{d\sigma_{L}(\gamma^{*}\rightarrow V)
\over d\sigma_{T}(\gamma^{*}\rightarrow V)}\,,
$
where $m_{V}$ is the mass of the vector meson.
The solid and dashed curves are for $W=15\,GeV$ and $W=150\,GeV$.

\item[Fig.~7]
~- The color dipole model predictions for the
dependence on the scaling variable
$Q^{2}+m_{V}^{2}$ of the polarization-unseparated
$d\sigma(\gamma^* \rightarrow V)/dt|_{t=0}=
d\sigma_{T}(\gamma^* \rightarrow V)/dt|_{t=0}+\epsilon
d\sigma_{L}(\gamma^* \rightarrow V)/dt|_{t=0}$ for
$\epsilon = 1$
at the HERA energy $W=100\,GeV$.

\item[Fig.~8]
~- The color dipole model
predictions for the $Q^2$ and $W$ dependence of the ratios
$\sigma(\gamma^{*}\rightarrow \rho'(2S))/
\sigma(\gamma^{*}\rightarrow \rho^{0})$ and
$\sigma(\gamma^{*}\rightarrow \phi'(2S))/
\sigma(\gamma^{*}\rightarrow \phi^{0})$
for the (T) and (L)
polarization of the vector mesons.

\item[Fig.~9]
~- The color dipole model
predictions for the $Q^{2}$ dependence of the ratio of
the polarization-unseparated forward production cross sections
$d\sigma(\gamma^{*}\rightarrow \rho'(2S))/
d\sigma(\gamma^{*}\rightarrow \rho^{0})$ and
$d\sigma(\gamma^{*}\rightarrow \phi'(2S))/
d\sigma(\gamma^{*}\rightarrow \phi^{0})$
for the polarization of
the virtual photon $\epsilon = 1$ at the HERA energy $W=100
\,GeV$.

\item[Fig.~10]
~- The color dipole model
predictions of the forward differential cross sections
$d\sigma_{L,T}(\gamma^* \rightarrow V')/dt|_{t=0}$ for
transversely(T)
(top boxes) and longitudinally (L)
(middle boxes) polarized radially excited
vector mesons $\rho'(2S)$ and
$\phi'(2S)$ and for the polarization-unseparated
$d\sigma(\gamma^* \rightarrow V')/dt|_{t=0}=
d\sigma_{T}(\gamma^* \rightarrow V')/dt|_{t=0}+\epsilon
d\sigma_{L}(\gamma^* \rightarrow V')/dt|_{t=0}$ for
$\epsilon = 1$ (bottom boxes)
as a function of the c.m.s. energy $W$
at different values of $Q^2$.

\item[Fig.~11]
~- The color dipole model
predictions for the energy dependence of the
ratio of the
polarization-unseparated forward production cross sections
$d\sigma(\gamma^{*}\rightarrow \phi^{0})/
d\sigma(\gamma^{*}\rightarrow \rho^{0})$
for the polarization of
the virtual photon $\epsilon = 1$  at different values of
$Q^{2}$.

\item[Fig.~12]
~- The color dipole model
predictions for the energy dependence of the
ratio of the
polarization-unseparated forward production cross sections
$d\sigma(\gamma^{*}\rightarrow J/\Psi)/
d\sigma(\gamma^{*}\rightarrow \rho^{0})$
for the polarization of
the virtual photon $\epsilon = 1$  at different values of
$Q^{2}$.

\item[Fig.~13]
~- Approximate scaling in the variable
$Q^{2}+m_{V}^{2}$ for the ratio of the
polarization-unseparated forward production cross sections
$d\sigma(\gamma^{*}\rightarrow \phi^{0})/
d\sigma(\gamma^{*}\rightarrow \rho^{0})$
and $d\sigma(\gamma^{*}\rightarrow J/\Psi)/
d\sigma(\gamma^{*}\rightarrow \rho^{0})$
for the polarization of
the virtual photon $\epsilon = 1$.
The horizontal dotted straight lines show the
ratio corresponding to Eq.~(\ref{eq:5.1}).

\item[Fig.~14]
~- Approximate scaling in the variable
$Q^{2}+m_{V}^{2}$ for the ratio of the
polarization-unseparated forward production cross sectionsf
$d\sigma(\gamma^{*}\rightarrow \phi^{0})/
d\sigma(\gamma^{*}\rightarrow \rho^{0})$
and $d\sigma(\gamma^{*}\rightarrow J/\Psi)/
d\sigma(\gamma^{*}\rightarrow \rho^{0})$.
at c.m.s. energy $W=150$\,GeV (the polarization of
the virtual photon $\epsilon = 1$).

\item[Fig.~15]
~- The $Q^{2}$ dependence of the coefficient functions
$g_{T,L}$ at $W=15\,GeV$ (dashed curve) and $W=150\,GeV$
(solid curve).

\item[Fig.~16]
~-
The dipole size dependence of the dipole cross section
extracted from the experimental data on photoproduction
and electroproduction of vector mesons:
the NMC data on $\phi^0$ and $\rho^0$ production \cite{NMCfirho},
the EMC data on $J/\Psi$ production \cite{EMCPsi,EMCPsiQ2},
the E687 data on $J/\Psi$ production \cite{E687},
the FNAL data on $\rho^0$ production \cite{Philownu},
the ZEUS data on $\phi^0$ production \cite{ZEUSphi},
the ZEUS data on $\rho^0$ production \cite{ZEUSrho94,ZEUSrho95,ZEUSrhoQ2},
the H1 data on $\rho^0$ production \cite{H1rho} and
the average of the
H1 and ZEUS data on $J/\Psi$ production \cite{H1Psi,ZEUSPsi}.
The dashed and solid curve show the dipole cross section
of the model \cite{NZHera,NNZscan} evaluated for
the c.m.s. energy $W=15$ and $W=70$ GeV respectively.
The data points at HERA energies and the corresponding solid curve
are multiplied by the factor 1.5.

\item[Fig.~17]
~-The $Q^{2}$ dependence of the parameter $\eta_{V}$ in the
representation (\ref{eq:6.6}) for the amplitude of leptoproduction
of different vector mesons at fixed energy $W=150$\,GeV.

\item[Fig.~18]
~-The $x_{eff}$ dependence of the parameter $\eta_{V}$ in the
representation (\ref{eq:6.6}) for the amplitude of leptoproduction
of different vector mesons at several values of $Q^{2}$.

~-
\pagebreak\\

\end{itemize}

\begin{table}[t]
\begin{center}
\begin{tabular}{||l||c|c|c|c|c|c|c|c|} \hline\hline
 parameter   & $\rho^{0}$ & $\rho'$ & $\phi^{0}$  & $\phi'$
             & $J/\Psi$   & $\Psi'$ & $\Upsilon$  & $\Upsilon'$ \\ \hline
$R^2$ [fm$^2$] &    1.37   &    1.39   &  0.69   & 0.83
             &    0.135   &    0.248   &  0.015   & 0.047       \\ \hline
$C$          &    0.36   &    0.28   &  0.53   & 0.44
             &    1.13   &    0.99   &  1.99   & 1.72       \\ \hline
$m_q$ [GeV]  &    0.15   &    0.15   &  0.30   & 0.30
             &    1.30   &    1.30   &  5.00   & 5.00       \\ \hline
$\alpha$     &            &    0.86   &          & 0.94
             &            &    1.20   &          & 1.53       \\ \hline
$R_{V}$ [fm] &    1.30   &    2.28   &  0.91   & 1.68
             &    0.41   &    0.83   &  0.19   & 0.42       \\ \hline
$\Delta m(2S-1S)$ [GeV] &      &    0.73   &   & 0.64
                     &      &    0.60   &   & 0.55  \\ \hline
$\Gamma( e^{+}e^{-})$ [keV]
             &    6.29   &    2.62   &  1.23   & 0.47
             &    5.06   &    1.78   &  1.20   & 0.54       \\ \hline
$\Gamma^{exp}(e^{+}e^{-})$ [keV]
             &    6.77   &       &  1.37   & 0.48
     &    5.36   &    2.14   &  1.34   & 0.56        \\
~~~~~~~
             & $\pm   0.32$   & $  $   & $\pm 0.05$
   & $\pm 0.14$
             &  $\pm  0.29$   & $\pm   0.21$   & $\pm 0.04$   &
$\pm 0.140$
\\ \hline\hline
\end{tabular}
\end{center}
\caption[.]{ \sl The parameters $R^2$, $C$, $m_q$
and $\alpha$
of the vector mesons wave function and some of the observables evaluated
with these wave functions: the r.m.s. $R_{V}$, the leptonic width
$\Gamma(e^{+}e^{-})$ and the $V'(2S)-V(1S)$ mass splitting.
The values of $\Gamma(e^{+}e^{-})$ from the Particle Data Tables
\cite{PDT} is shown for the comparison.}
\end{table}

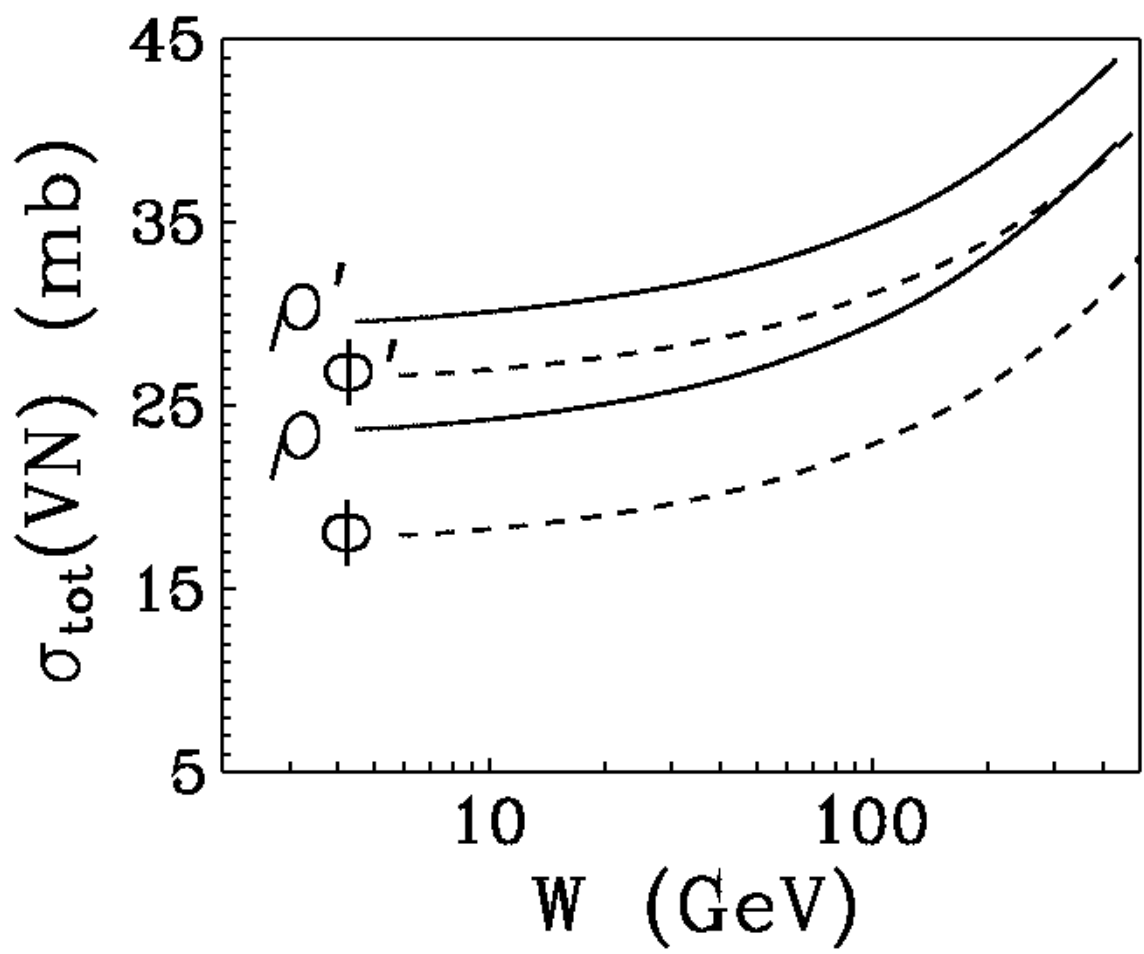
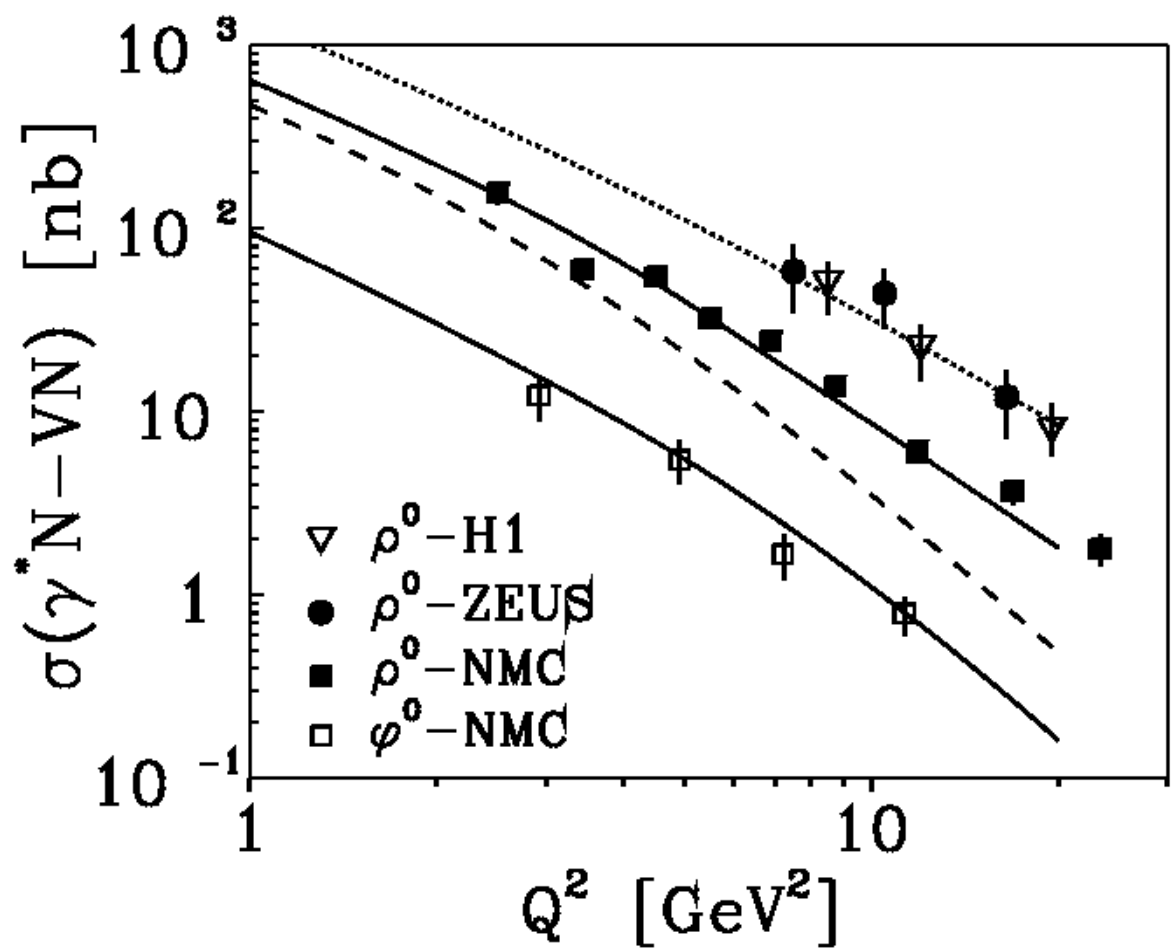
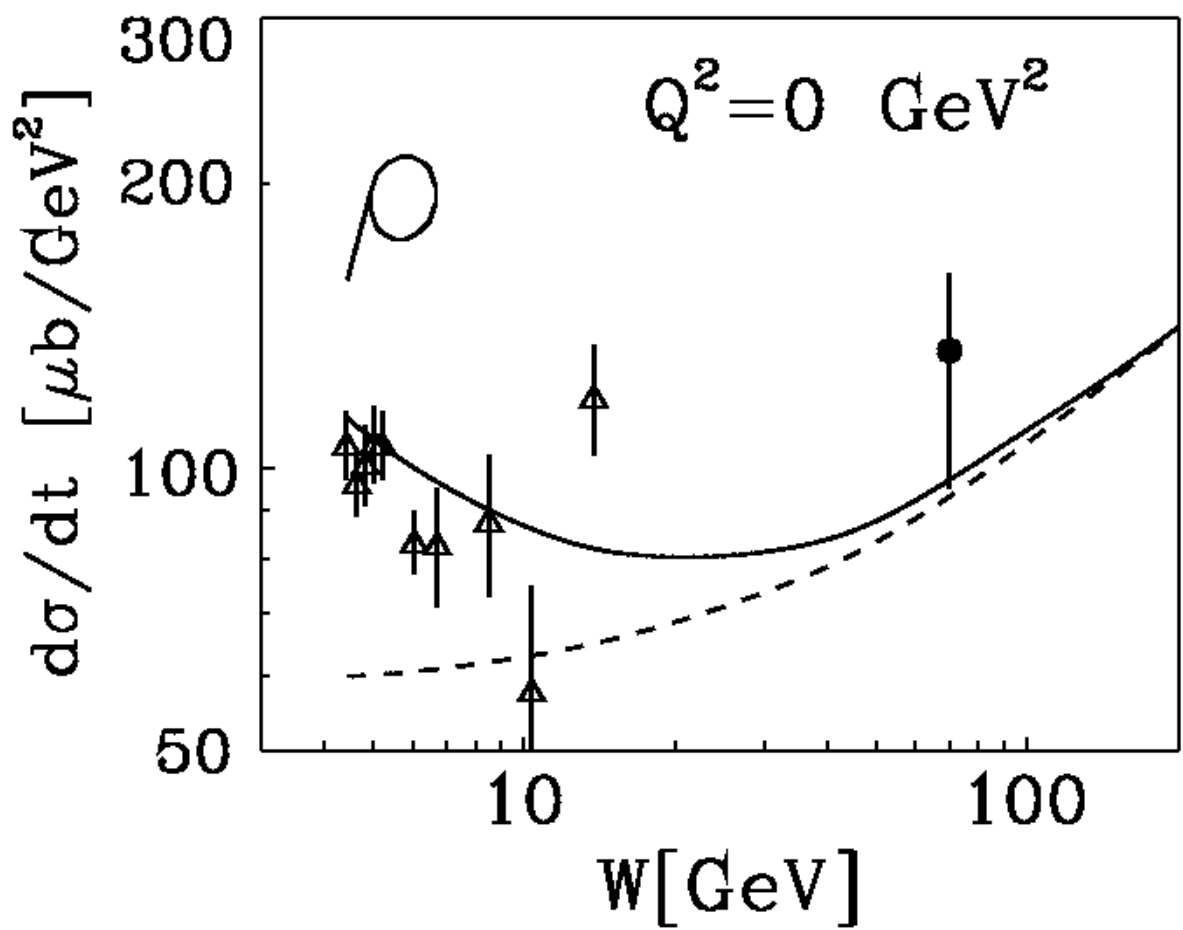
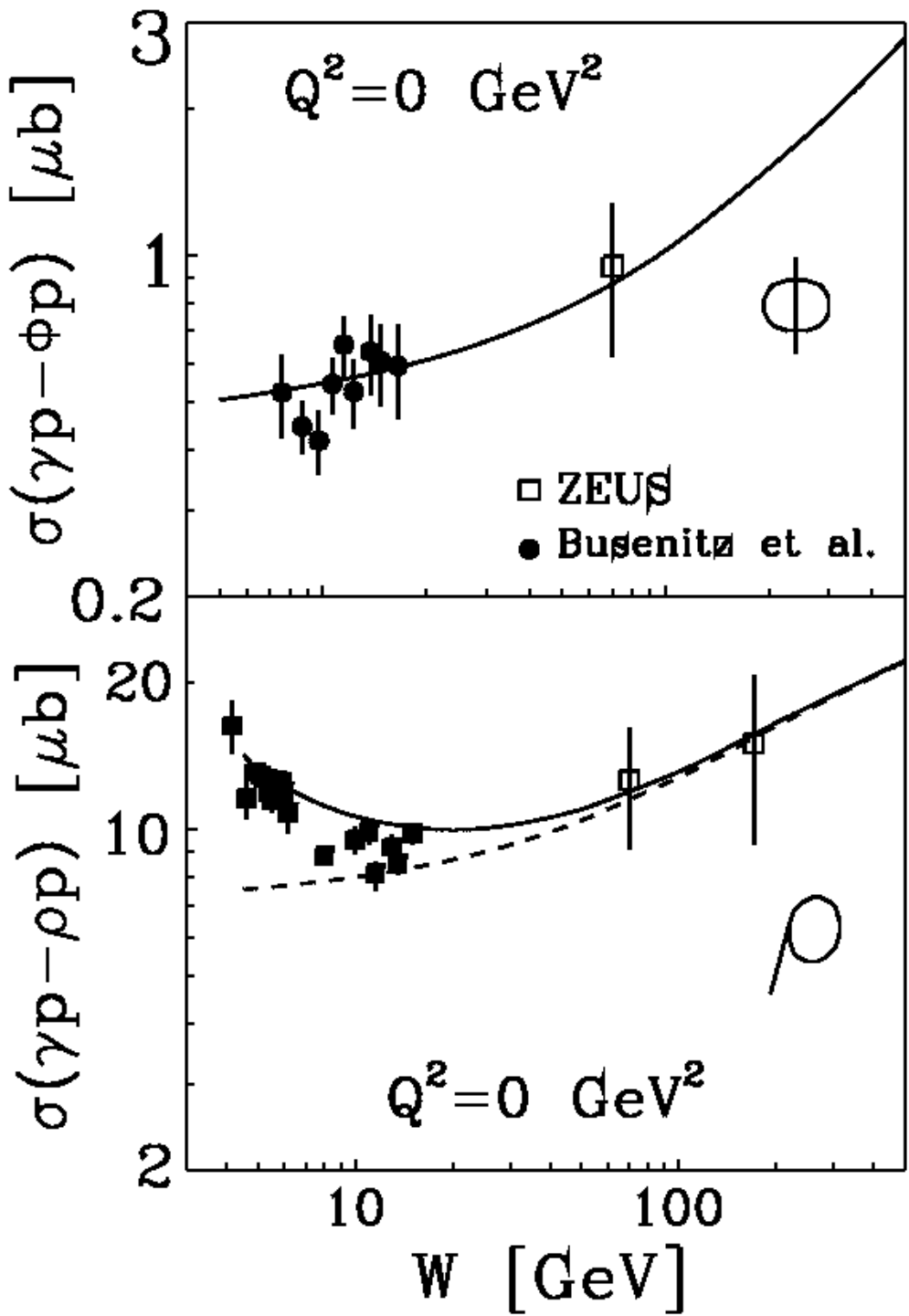
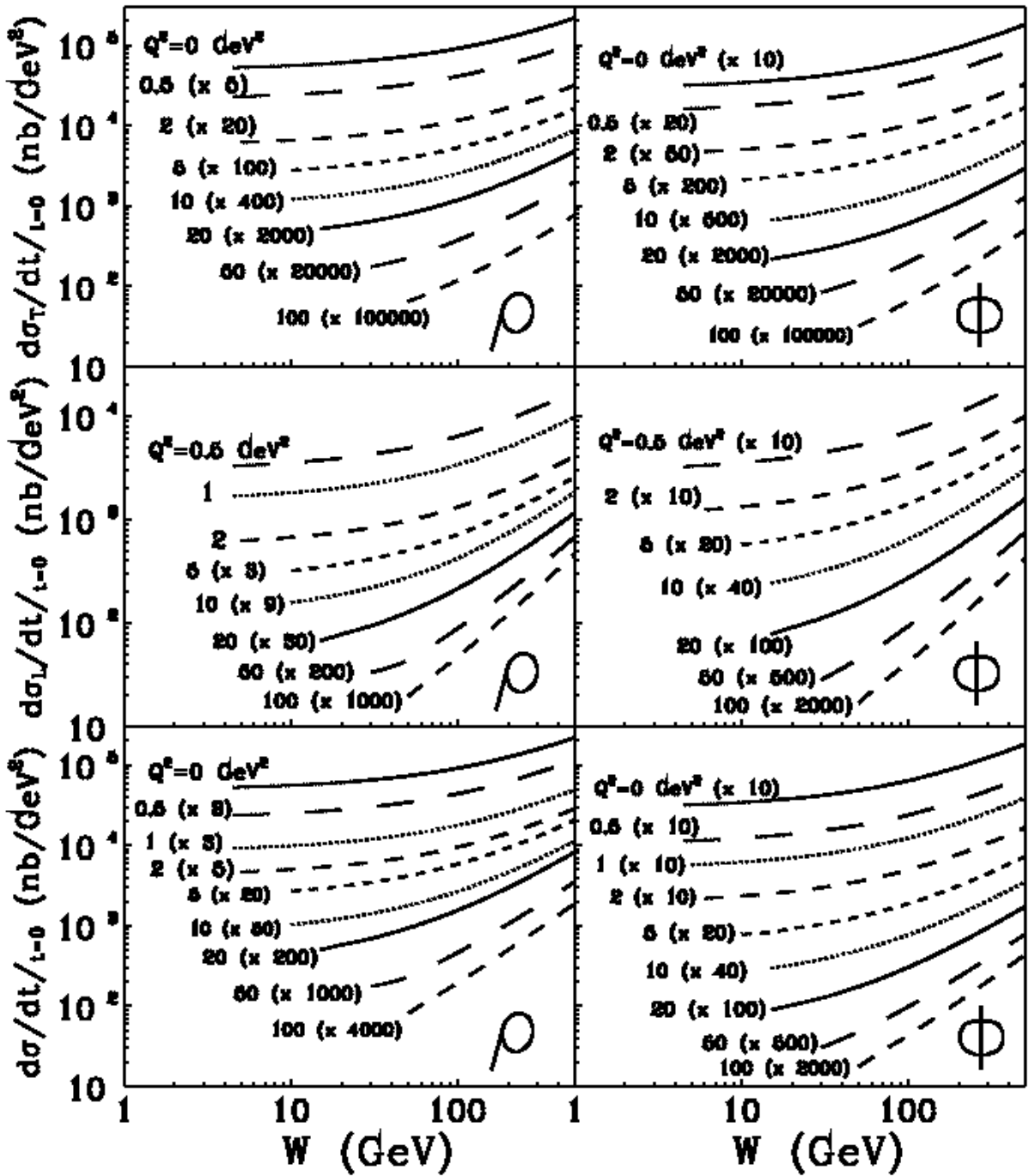
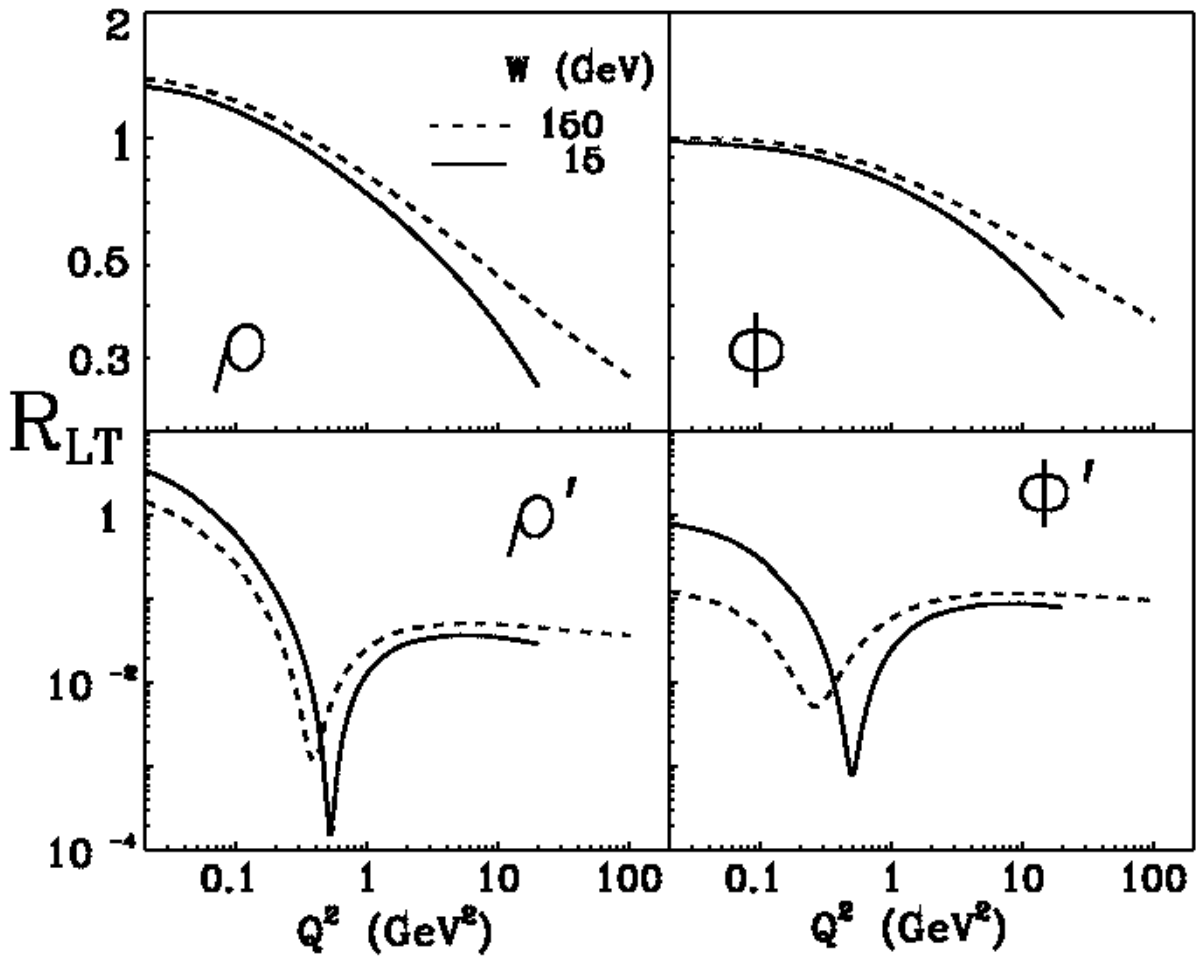
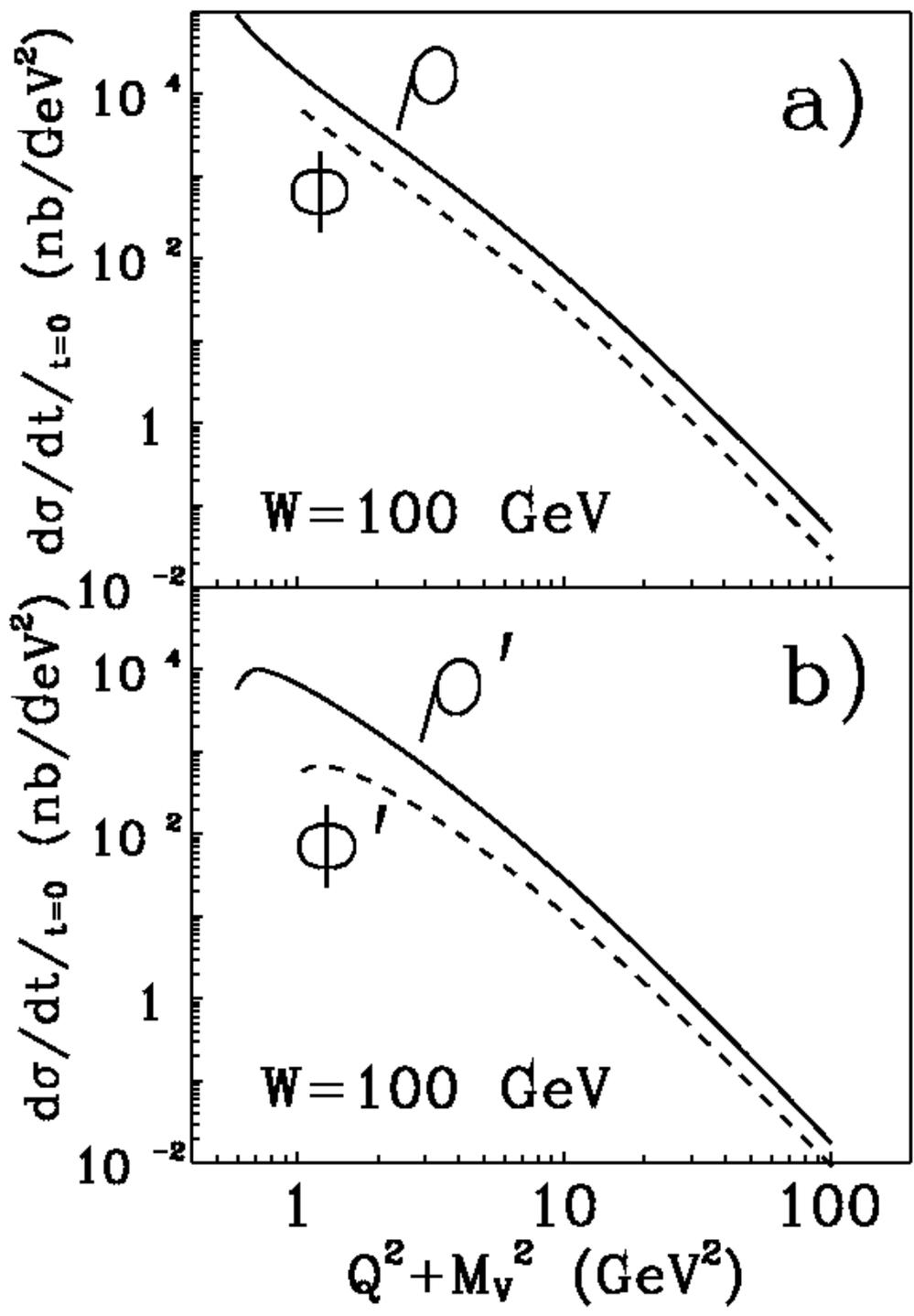
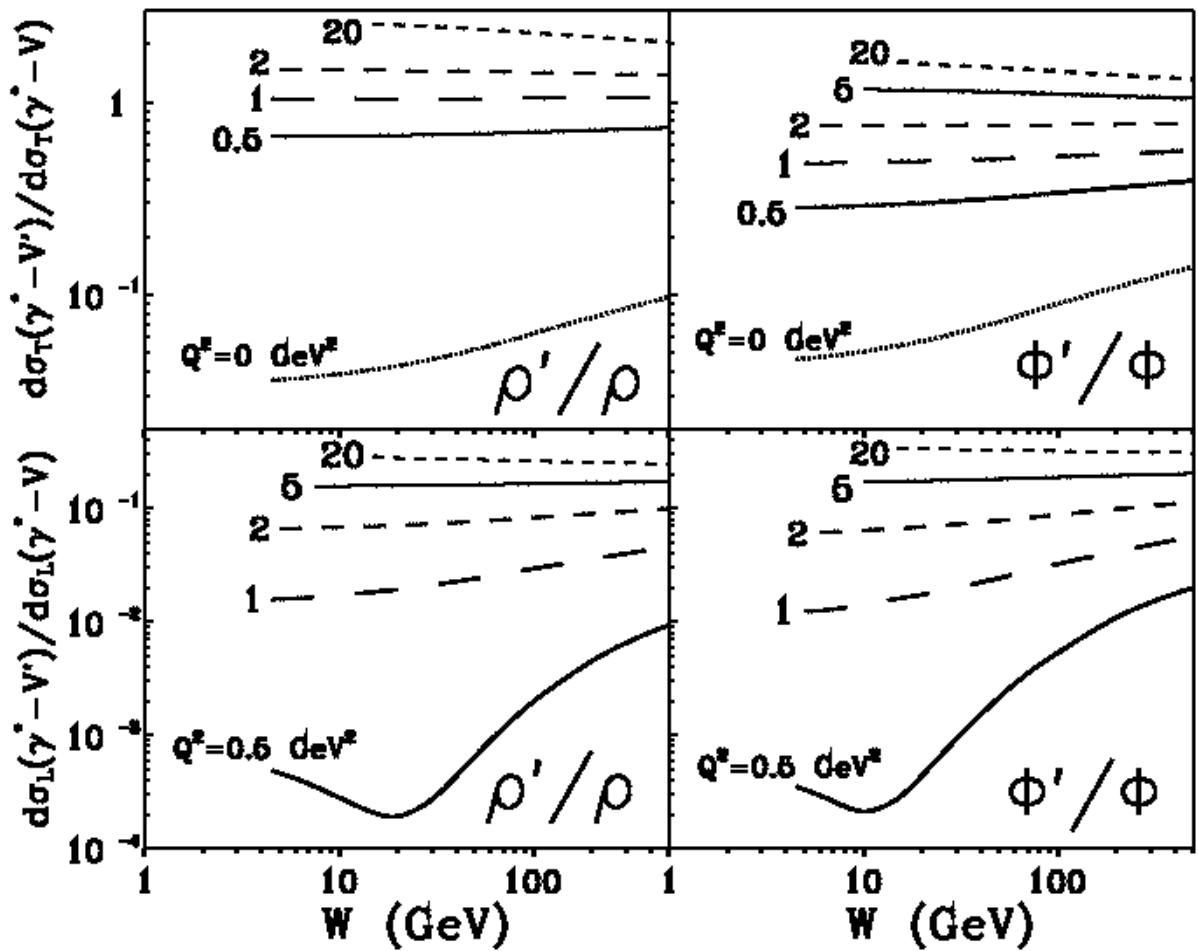
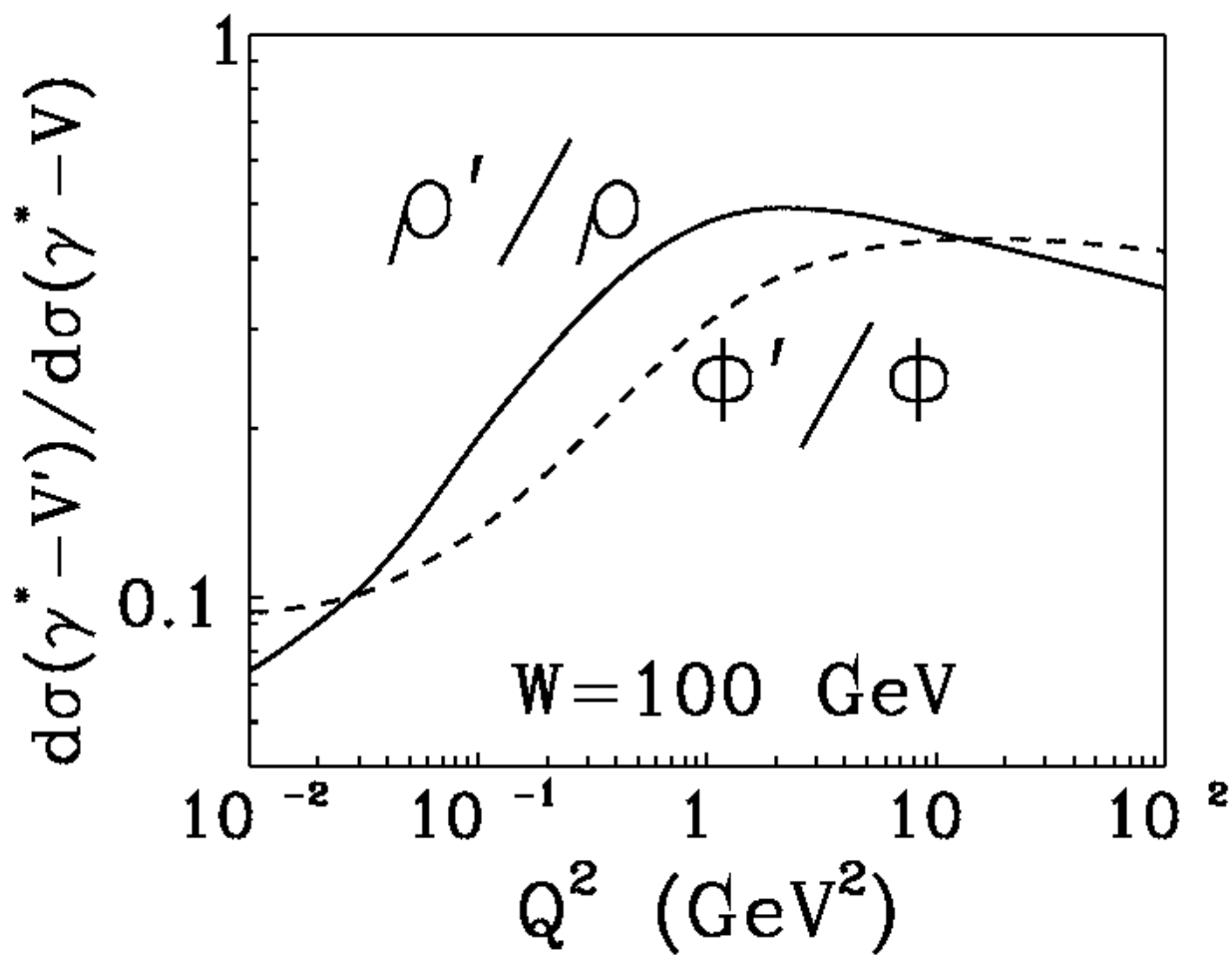
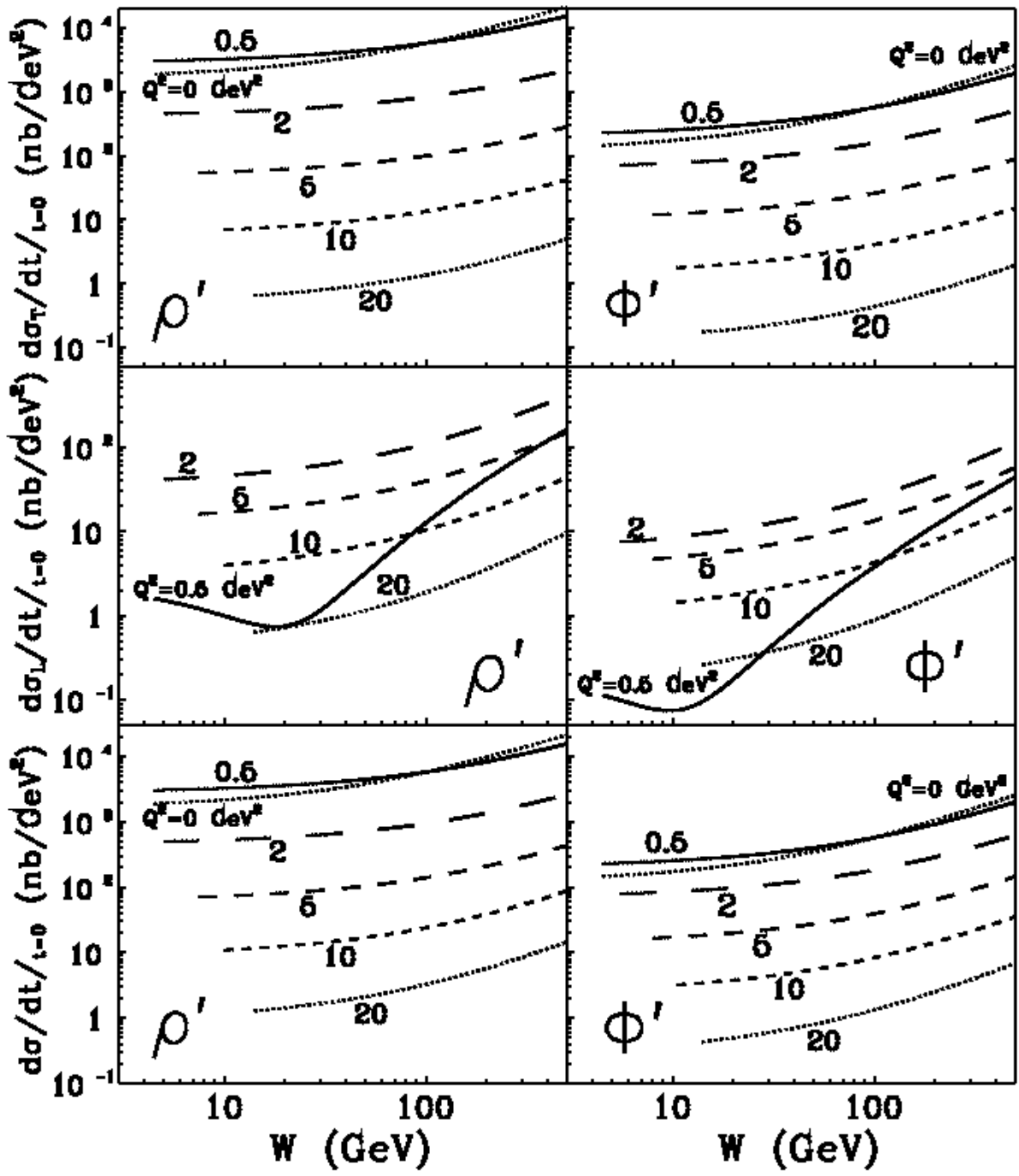
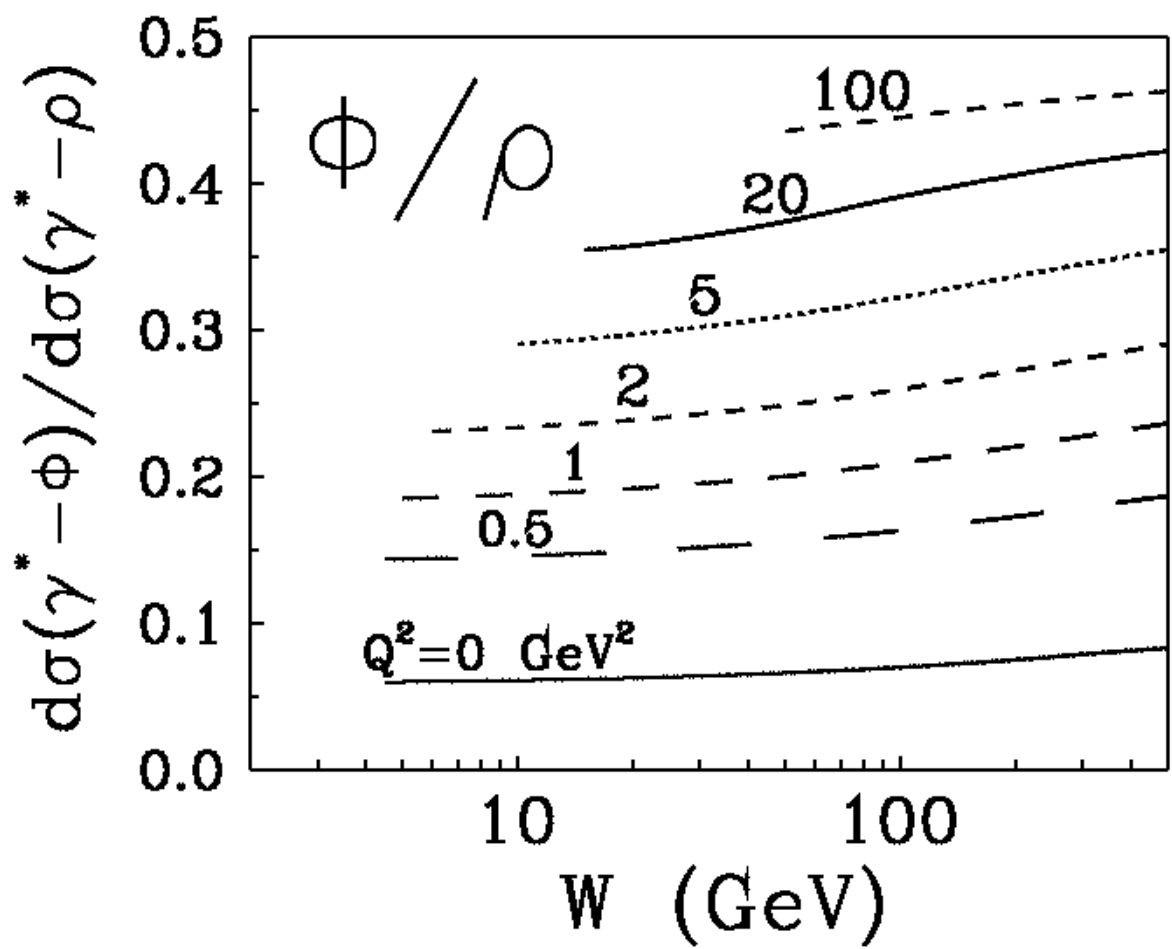
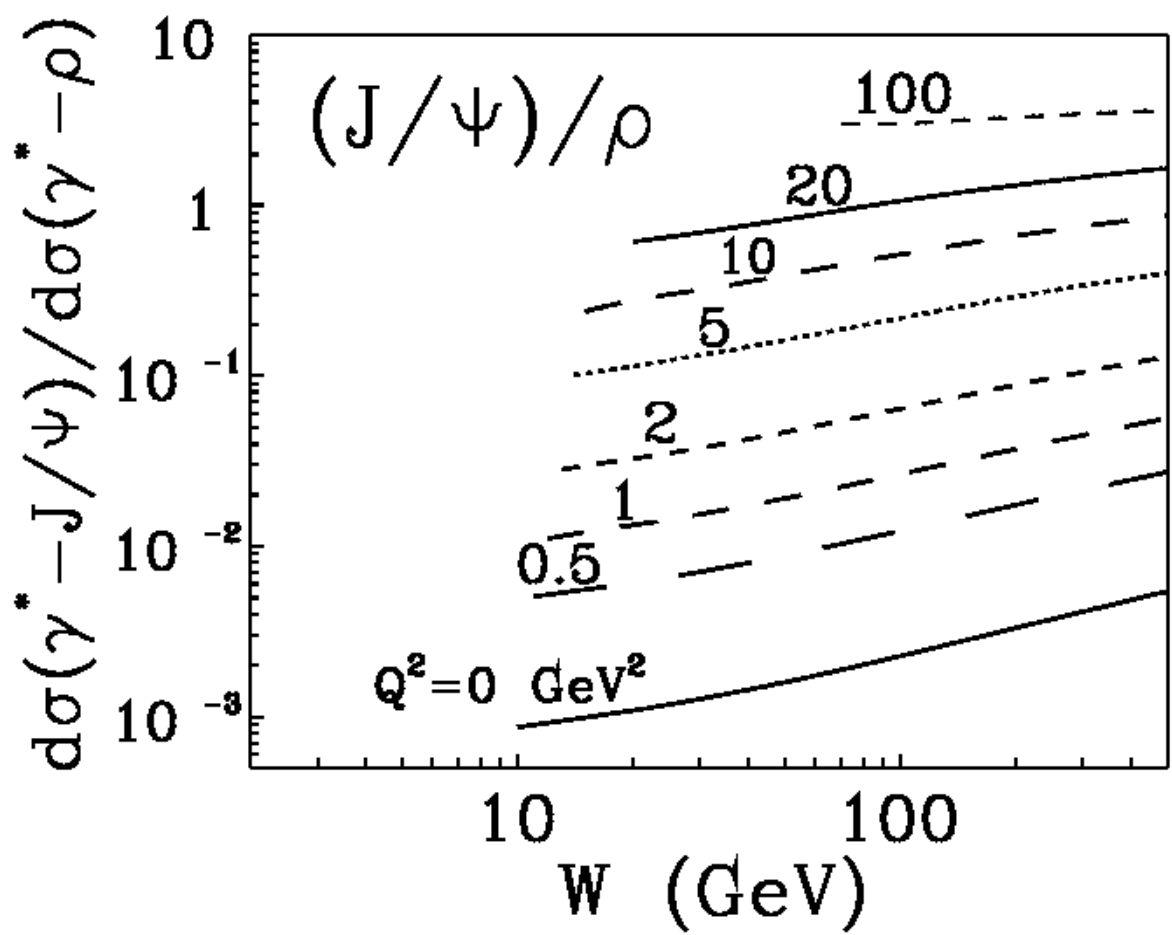
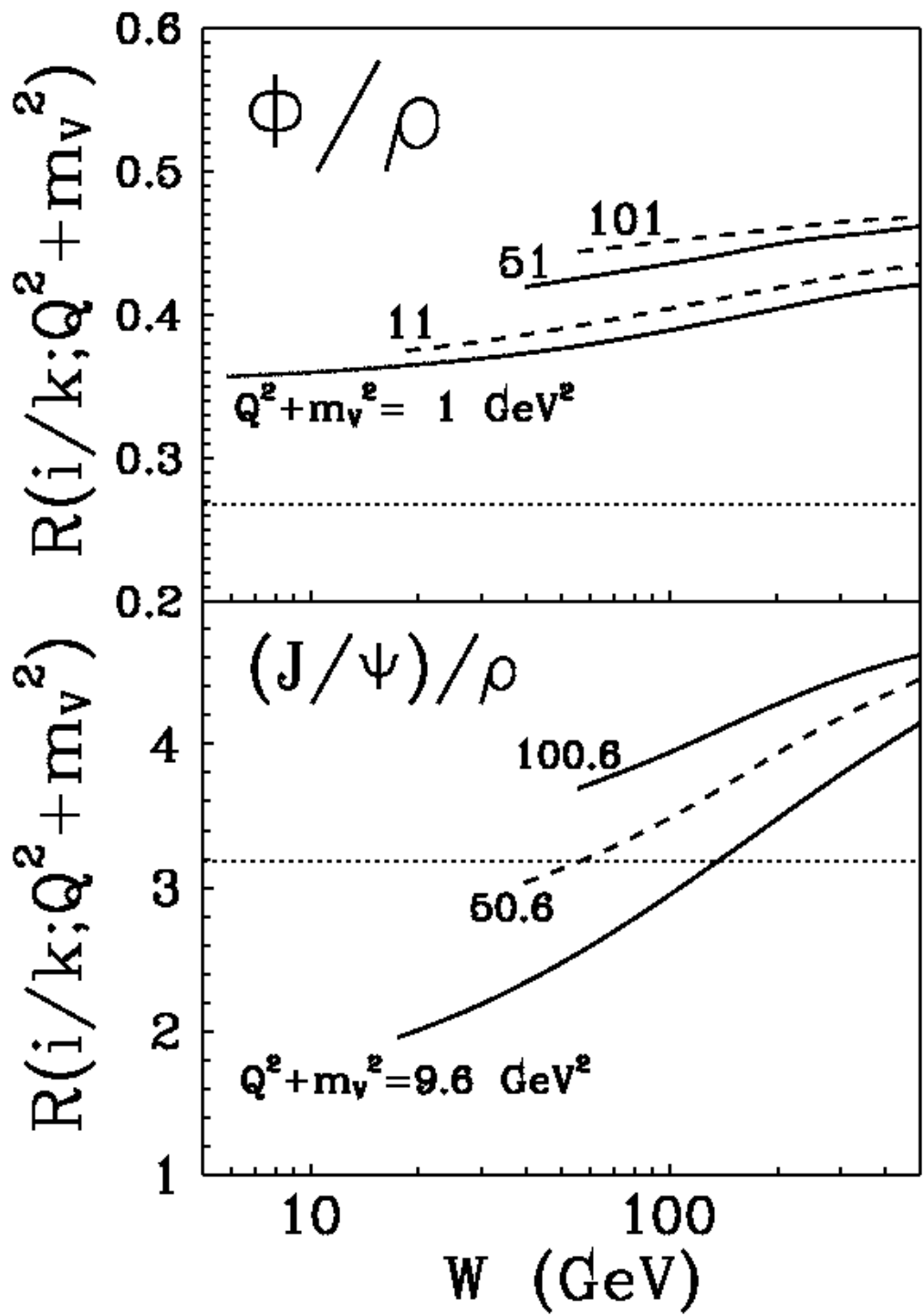
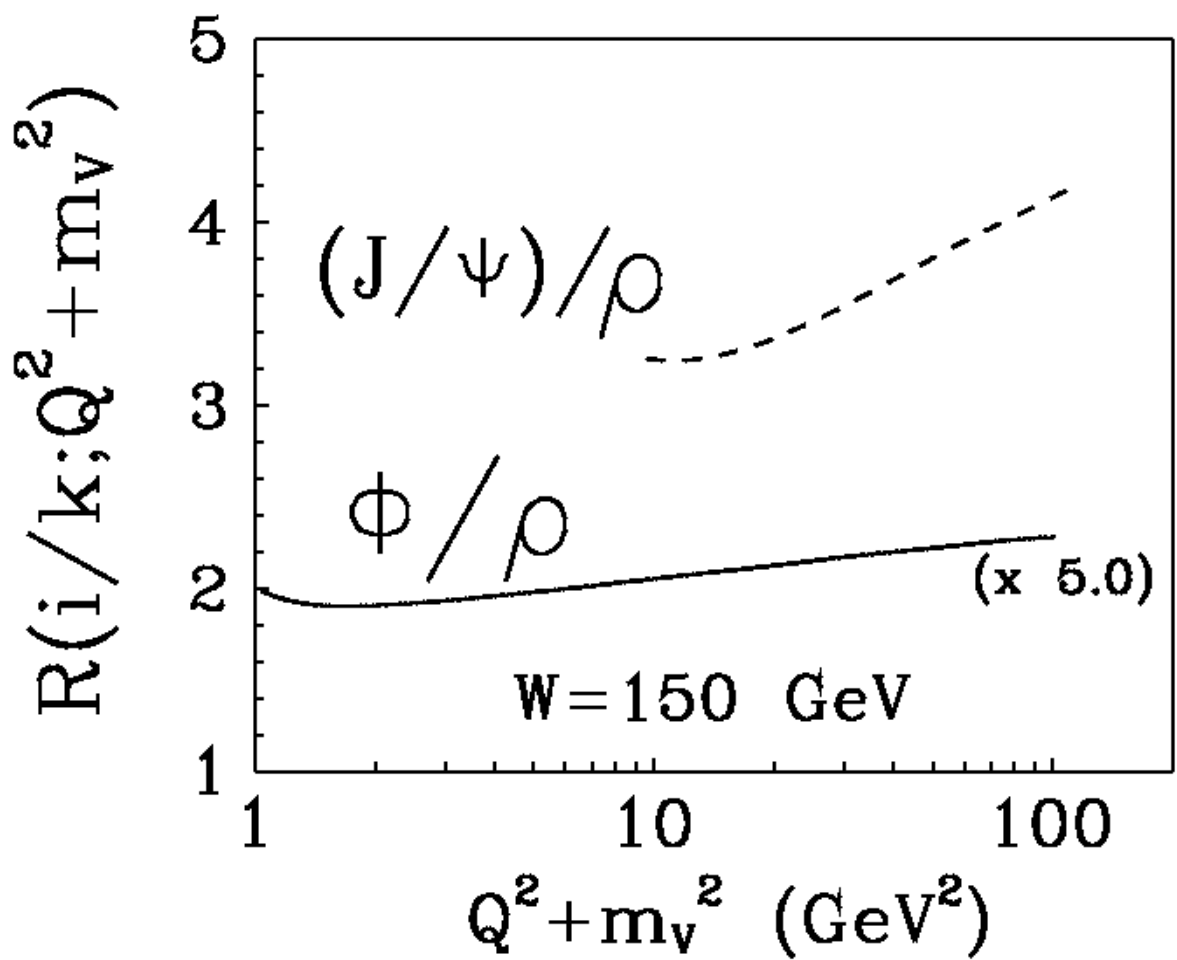
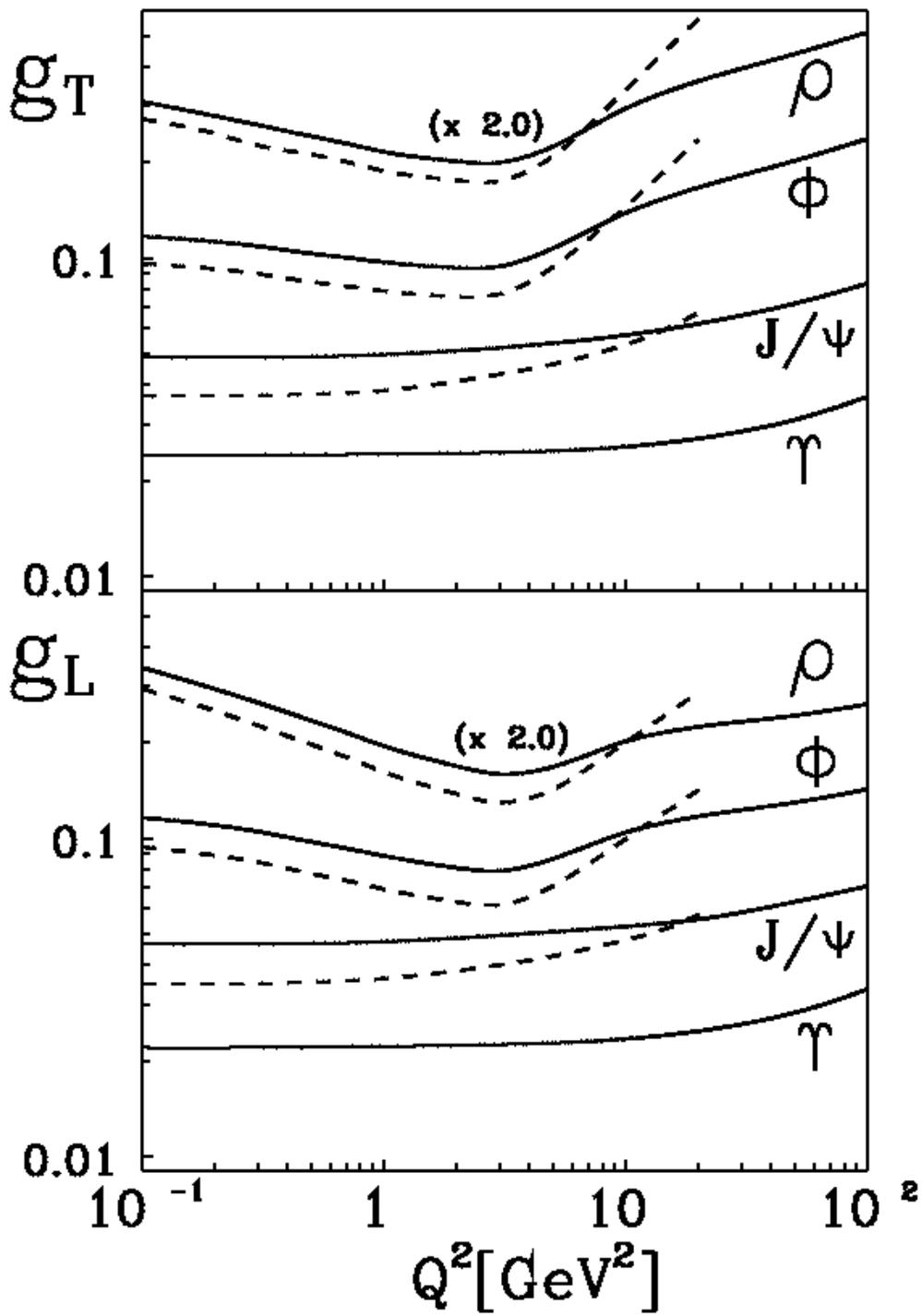
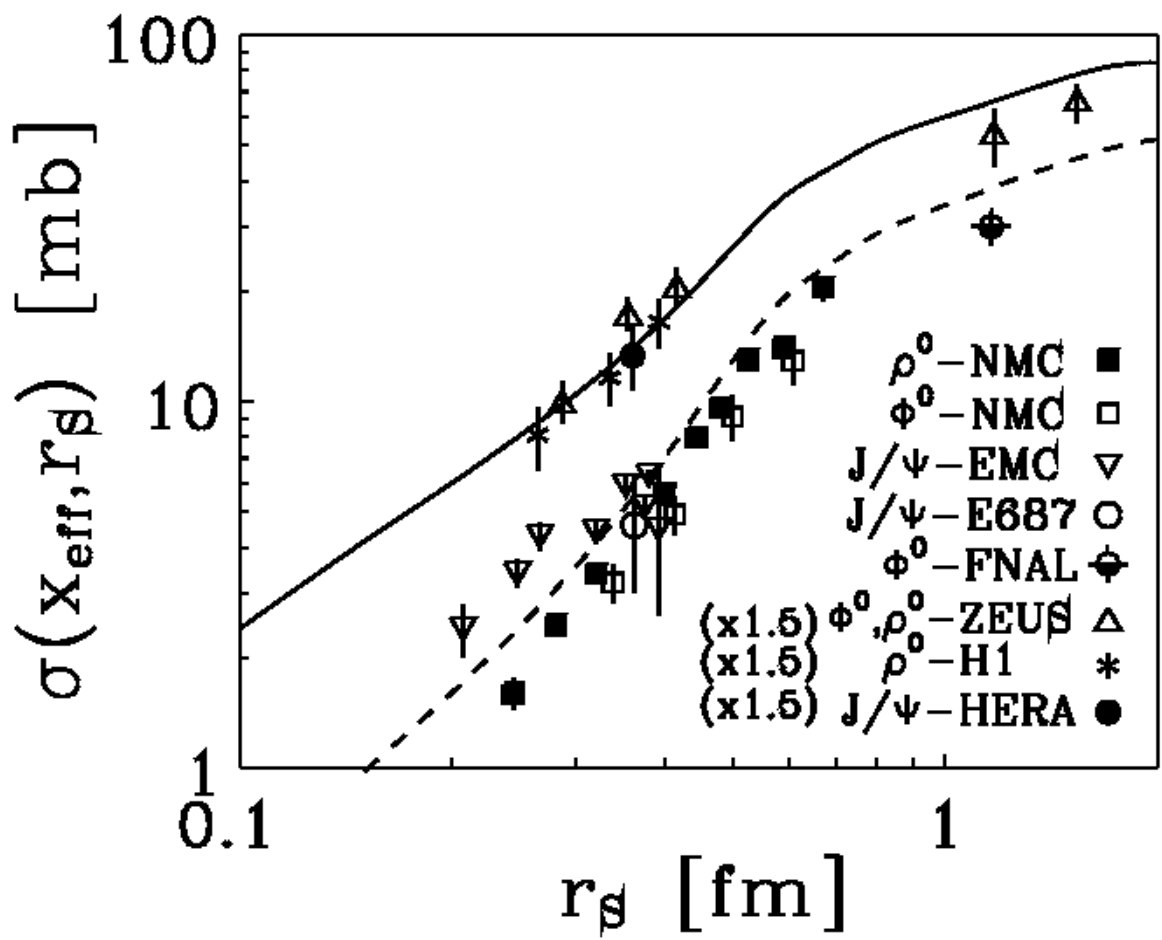
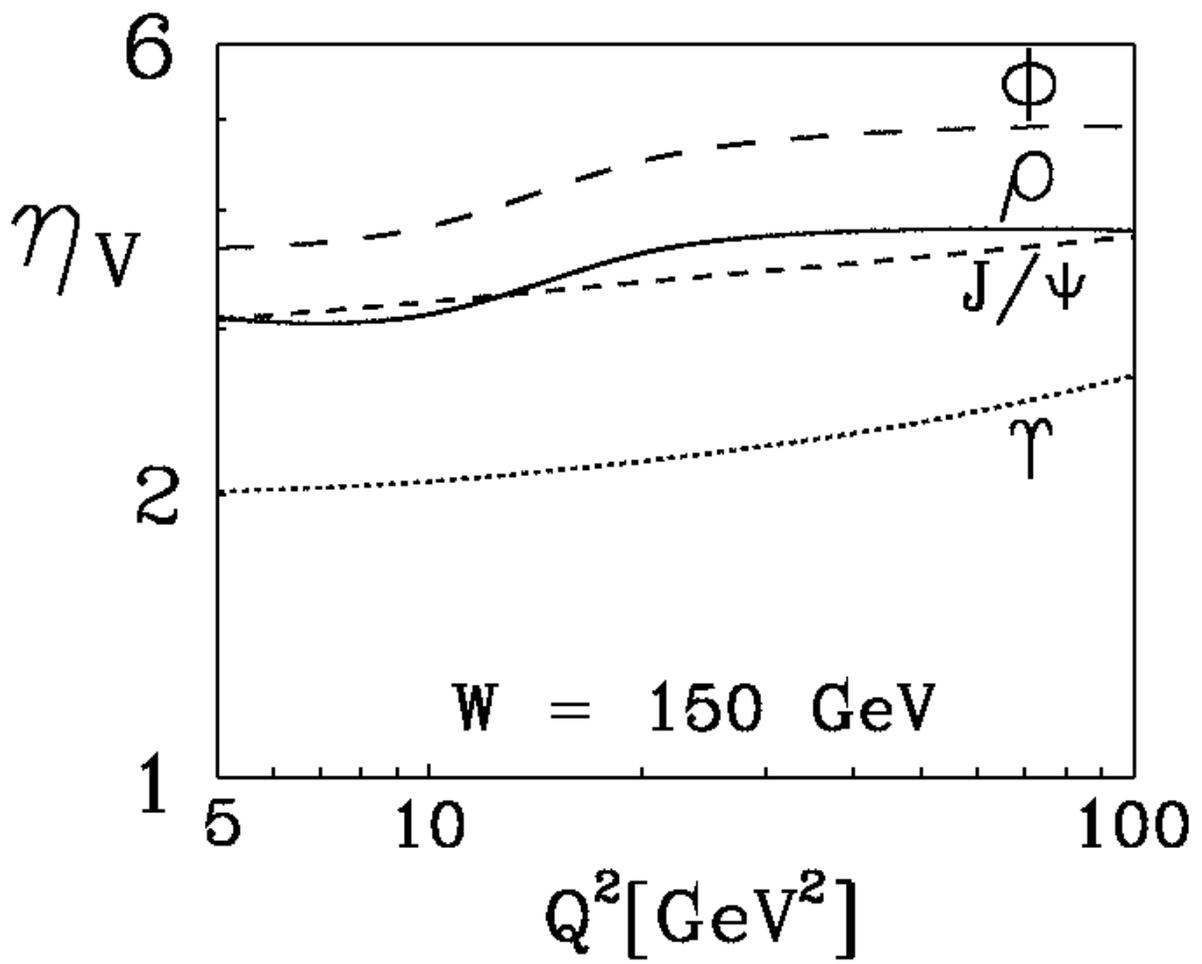
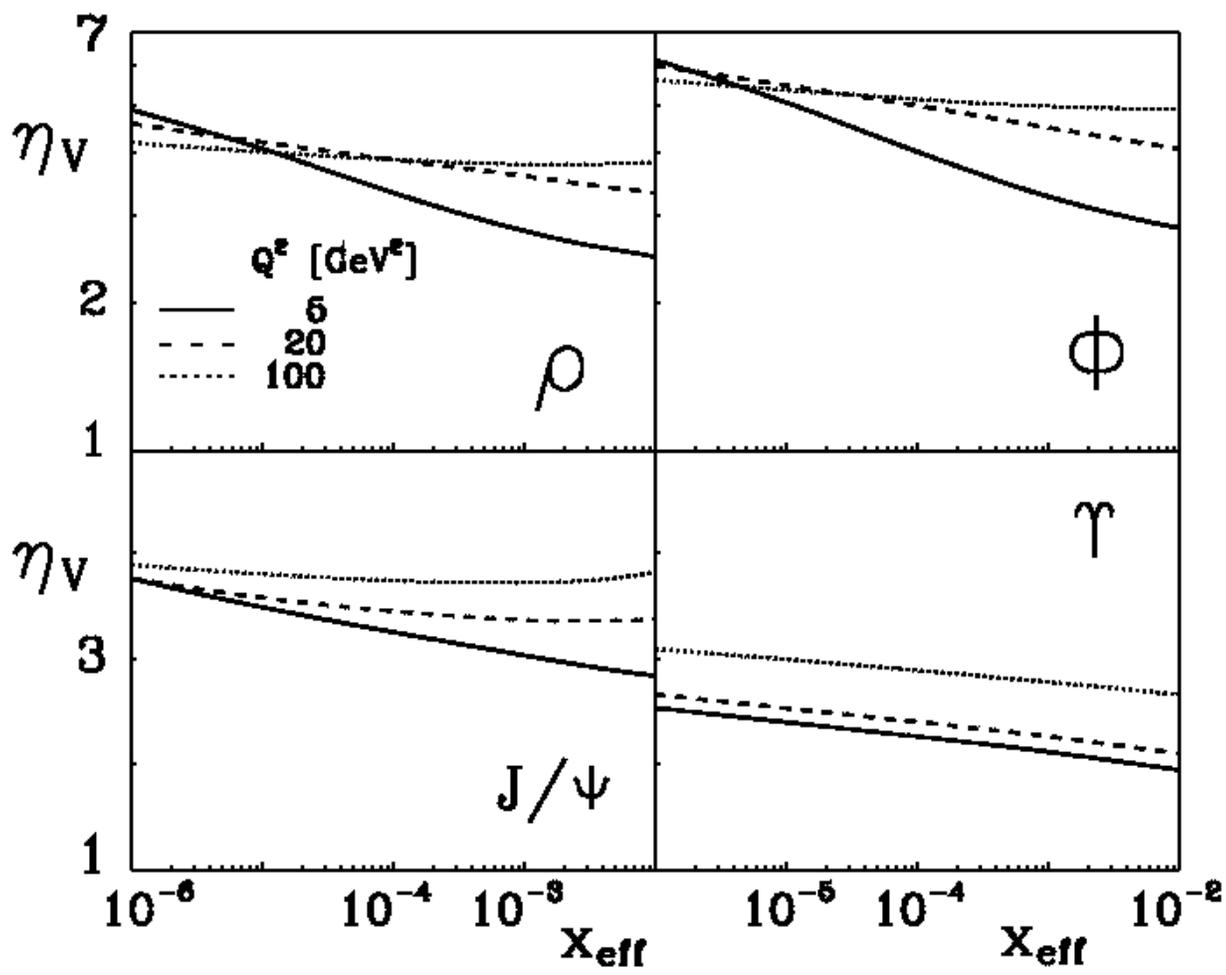
\end{document}